\documentclass[journal]{IEEEtran}   
\usepackage{multirow}
\usepackage{diagbox}
\usepackage{amssymb}
\usepackage{amsmath}
\usepackage{graphicx}
\usepackage{cite}
\usepackage{citesort}
\usepackage{subfigure}
\usepackage{graphicx,epstopdf}
\usepackage{epsfig}	
\usepackage{cite,graphicx,amsmath,amssymb}
\usepackage{comment}

\usepackage{amssymb}
\usepackage{amsmath}
\usepackage{cite}
\usepackage{url}
\usepackage{xcolor}
\usepackage{cite,graphicx,amsmath,amssymb}
\usepackage{subfigure}
\usepackage{citesort}
\usepackage{fancyhdr}
\usepackage{mdwmath}
\usepackage{mdwtab}
\usepackage{caption}
\usepackage{amsthm}
\usepackage{lipsum}

\newtheorem{theorem}{Theorem}

\newtheorem{lemma}{Lemma}

\newtheorem{corollary}{Corollary}

\newtheorem{remark}{Remark}  

\usepackage [latin1]{inputenc}

\makeatletter
\def\ScaleIfNeeded{%
\ifdim\Gin@nat@width>\linewidth \linewidth \else \Gin@nat@width
\fi } \makeatother

\begin{document}

\title{\Huge{Performance Analysis of Intelligent Reflecting Surface Assisted NOMA Networks}}

\author{ Xinwei~Yue,~\IEEEmembership{Member,~IEEE} and Yuanwei\ Liu,~\IEEEmembership{Senior~IEEE}

\thanks{X. Yue is with the School of Information and Communication Engineering and also with the Key Laboratory of Modern Measurement $\&$ Control Technology, Ministry of Education and also, Beijing Information Science and Technology University, Beijing 100101, China (email: xinwei.yue@bistu.edu.cn).}
\thanks{Y. Liu is with the School of Electronic Engineering and Computer Science, Queen Mary University of London, London E1 4NS, U.K. (email: yuanwei.liu@qmul.ac.uk).}
}

\maketitle

\begin{abstract}
Intelligent reflecting surface (IRS) is a promising technology to enhance the coverage and performance of wireless networks. We consider the application of IRS to non-orthogonal multiple access (NOMA), where a base station transmits superposed signals to multiple users by the virtue of an IRS. The performance of an IRS-assisted NOMA networks with imperfect successive interference cancellation (ipSIC) and perfect successive interference cancellation (pSIC) is investigated by invoking 1-bit coding scheme. In particular, we derive new exact and asymptotic expressions for both outage probability and ergodic rate of the $m$-th user with ipSIC/pSIC. Based on analytical results, the diversity order of the $m$-th user with pSIC is in connection with the number of reflecting elements and channel ordering. The high signal-to-noise radio (SNR) slope of ergodic rate for the $m$-th user is obtained. The throughput and energy efficiency of  IRS-NOMA networks are discussed both in delay-limited and delay-tolerant transmission modes. Additionally, we derive new exact expressions of outage probability and ergodic rate for IRS-assisted orthogonal multiple access (IRS-OMA). Numerical results are presented to substantiate our analyses and demonstrate that: i) The outage behaviors of IRS-NOMA are superior to that of IRS-OMA and relaying schemes; ii) The $M$-th user has a larger ergodic rate than IRS-OMA and benchmarks. However, the ergodic performance of the $m$-th user exceeds relaying schemes in the low SNR regime; and iii) The IRS-assisted NOMA networks have ability to achieve the enhanced energy efficiency compared to conventional cooperative communications.
\end{abstract}

\section{Introduction}
With the evolution of wireless communication networks, the fifth-generation (5G) and beyond has sparked a lot of concerns on high data rate, massive connectivity and spectrum utilization. The standards of 5G new radio have been completed currently and researchers are exploring the potential of emerging technologies for the next-generation communications \cite{Liu8114722Beyond,Saad6G,Zhang8766143}. As one of a promising multiple access candidate, non-orthogonal multiple access (NOMA) has the advantages in terms of spectral efficiency and link density. The distinctive feature of NOMA is that multiple users are allowed to occupy the same time/bandwidth resource blocks by utilizing the superposition coding scheme \cite{Tse2005,Ding2017Application}. It has been demonstrated that NOMA has ability to attain the better outage probability and ergodic rate compared to conventional orthogonal multiple access (OMA) \cite{Ding2014performance,Yue8370069Unified}. For the emerging sixth-generation (6G) communication networks, it becomes pivotal to support massive intelligent equipments with different  requirements.

By extending NOMA to cooperative communications, cooperative NOMA was proposed in \cite{Ding2014Cooperative}, where the nearby user with better channel condition was referred to half-duplex (HD) decode-and-forward (DF) relaying to transfer the signals for the distant users. To further enhance spectrum efficiency, the authors of \cite{Zhong7572025,Yue8026173} investigated the outage behavior and ergodic rate of full-duplex (FD) cooperative NOMA, where the performance of FD NOMA outperforms HD NOMA in the low signal-to-noise radio (SNR). With the emphasis on green communication, the simultaneous wireless power transfer (SWIPT) based NOMA system was studied in \cite{Liu7445146SWIPT}, where the nearby user is viewed as DF relaying to forward the information. On the other hand, the authors of \cite{Liang7876764AF} analyzed the outage performance of a pair of users for amplify-and-forward (AF) relaying based NOMA systems. As a further development, the outage probability and ergodic performance of multiple users for AF NOMA systems were surveyed in \cite{Men2017,Yue7812773FixedAF} over Nakagami-$m$ fading channels. Explicit insights for understanding the impact of FD mode on AF NOMA system, the authors of \cite{Liau8AFFDNOMA} characterized the outage behaviors of users with an opportunistic power split factor. Apart from the above works, NOMA technique has been widely applied to multiple communication scenarios. Regarding to safety applications, the secrecy outage probability of a pair of users was analyzed in \cite{Yue2019UnifiedPLS} for NOMA networks by invoking stochastic geometry. With the objective of improving terrestrial user connections, the authors of \cite{Liu8641425} highlighted the trajectory design and power allocation of NOMA-based unmanned aerial vehicle networks. Additionally, the application of NOMA to satellite communications was investigated \cite{Yan8374960Land}, where the ergodic capacity and energy efficiency are derived analytically.

In view of recent attentions, intelligent reflecting surface (IRS) has been as a prospective technology for 6G wireless communications \cite{ZhaoIRS6G}. More specifically, IRS is a low-cost planar array consisting of a large of passive reflecting elements, which is ability to reconfigure the wireless propagation environment through a programmable controller. In \cite{Cui2014}, the authors have proposed the concept of digital metamaterials, which manipulates the electromagnetic waves by coding `0' and `1' elements with control sequences (i.e., 1-bit coding). Recently, several application deployment scenarios of IRS-aided were introduced \cite{WuTowards2019,Liu2020Opportunities} that: 1) Creating the line-of-sight (LoS) link between the BS and users via signal reflection; 2) Applying IRS to enhance the physical layer secrecy; 3) IRS-assisted to realize SWIPT for a large amount of devices; and 4) Deploying IRS on indoor walls, mall or stadium to provide the hot spot services and so on. The differences between IRS and active relaying such as DF relaying are mainly that IRS does not make use of any active transmitting components but only reflects the received signals to destination node through a passive array. Except these, IRS does not requires self-interference cancellation operations relative to full-duplex relaying which are costly to implement. Compared to DF relaying, the authors in \cite{Emil2019} revealed that IRS has the ability to attain the higher energy efficiency on the condition of the required large rates. The achievable rate performance of IRS-assisted transmission outperformed that of DF relaying \cite{Renzo2020IRS}, when the size of the IRS is sufficiently larger.  As a further advance,  the authors surveyed the symbol error probability of IRS-aided wireless communication networks in a general mathematical framework \cite{Basar8796365}. When the eavesdropping channels were stronger than that of legitimate channels, the authors of \cite{Cui8723525PLAviaIRS} maximized the secrecy rate of legitimate users by designing the IRS's reflecting beamforming. In \cite{Huang8741198}, the energy efficient designs of IRS-based wireless communications were developed to guarantee individual link budget for users. Furthermore, the security performance of IRS-aided SWIPT was surveyed in \cite{Xiong9133120}, where  the energy efficiency is maximized by combining the artificial noise, transmitting beamforming and phase shifts of IRS. For indoor scenarios, the authors of \cite{Renzo2020Indoor} maximized the channel capacity of indoor millimeter-wave by exploiting IRS's reflection elements.

In light of the above discussions, researchers have begun to study the co-existence of IRS and NOMA \cite{ZhengIRSUserpairing,Fu2019,Yang2019IRS}.  Given the users' rate, the authors in \cite{ZhengIRSUserpairing} analyzed the IRS reflection with discrete phase shifts for IRS-aided NOMA and OMA. On the condition of user ordering, the beamforming vectors and phase shift matrix were jointly optimized to reduce the transmitting power \cite{Fu2019}. Furthermore, the authors of \cite{Yang2019IRS} maximized the minimum the achievable rates of users to ensure user's fairness and improve the rate performance. In \cite{Zuo2020}, the maximization of system throughput was formulated  for IRS-NOMA by taking decoding order and reflection coefficients into consideration. For multi-antenna scenarios, the system sum rate of IRS-NOMA was improved by exploiting the fixed reflecting elements in \cite{Mu2019IRSmultiantennas}. To overcome the hardware limitations, a simple transmission of IRS-NOMA was designed in \cite{Ding2019IRS}, where the outage probability of single user is derived with on-off control. As a further potential enhancement, the authors of \cite{Ding2019Shifting} further studied the impact of coherent phase shifting and random phase shifting on outage behaviors for IRS-NOMA networks. In \cite{Hou2019IRS}, the outage probability and ergodic rate of nearby user for IRS-NOMA were examined by designing the passive beamforming weights of IRS. Additionally, the secrecy outage performance of IRS-NOMA was evaluated in \cite{Yuan2020IRS}, where the use of IRS improves the security behaviors compared to conventional NOMA.

\subsection{Motivation and Contributions}
As stated from conventional NOMA literatures that the decoding order of SIC-based at the users is mainly
decided by the users' channel power gains. However the directions of users' channel can be manipulated by adjusting the IRS's reflection signal amplitudes or phase shifts in IRS-assisted NOMA networks, where the combination of IRS and NOMA is capable of enhancing both the spectrum and energy efficiency of communication systems simultaneously. In \cite{Cui2014}, the digital metamaterial has ability to manipulate electromagnetic waves by programming different coding sequences. Consequently, research on using the thought of 1-bit coding scheme to analyze the performance of wireless communication systems is still imperative. Inspired by this treatise, we specifically investigate the performance of IRS-assisted NOMA with the aid of the thought of 1-bit coding, where an IRS can be regarded as a relay forwarding the information to multiple NOMA users.
In addition, there are undesirable factors i.e., error propagation and quantization error in the SIC process for practical scenarios, which will result in decoding errors. It is significant to take the residual interference from SIC procedure into consideration. Hence we focus our attention on discussing the impact of residual interference from  imperfect successive interference cancellation (ipSIC) on outage probability, ergodic rate and energy efficiency for IRS-NOMA networks. Additionally, the outage probability and ergodic rate of IRS-OMA with 1-bit coding are surveyed carefully. According to the above explanations, the primary contributions of this paper are summarized as follows:
\begin{enumerate}
  \item  We derive exact and asymptotic expressions of outage probability for the $m$-th user with ipSIC and perfect successive interference cancellation (pSIC) in IRS-NOMA networks. Based on theoretical analyses, the diversity order for IRS-NOMA is obtained. We demonstrate that the diversity orders of the $m$-th user with ipSIC/pSIC are in connection with the number of reflecting elements of IRS and channel ordering. We also derive the closed-form expression of outage probability for IRS-assisted OMA.
  \item We confirm that the outage behaviors of IRS-NOMA are superior to that of IRS-OMA, AF relaying and FD/HD DF relaying. Furthermore, we investigate the impact of IRS's deployment on the outage behaviors of IRS-NOMA networks. We observe that when the IRS is deployed closely to the BS, the enhanced outage performance is achieved. As the IRS departs from BS, the LoS deteriorates and the outage probability increases.
  \item  We derive the exact expressions of ergodic rate of the $m$-th user for IRS-NOMA networks. To obtain further more insights, we derive exact expression of ergodic rate for the $m$-th user and obtain the high SNR slopes. We observe that the ergodic rate of the $m$-th user converges to a throughput ceiling in the high SNR regime. As the number of reflecting elements increase, the ergodic performance of the $M$-th user is becoming higher compared to relaying schemes. We also derive exact expression of ergodic rate for IRS-OMA.
  \item  We study the throughput and energy efficiency of IRS-NOMA networks in both delay-limited and delay-tolerant transmission modes. For delay-limited transmission mode, the energy efficiency of IRS-NOMA outperforms that of IRS-OMA and converge to a constant value at high SNRs, while the IRS-NOMA and IRS-OMA achieve the larger energy efficiency than AF relaying and FD/HD DF relaying. In delay-tolerant transmission mode, the energy efficiency of IRS-NOMA is much larger than that of these benchmarks.
\end{enumerate}

\subsection{Organization and Notations}
The rest of this paper is organized as follows. In Section \ref{System Model}, the network model and transmission formulation are introduced in detail. In Section \ref{Outage probability}, new exact expressions of outage probability for IRS-NOMA are derived. Then the ergodic rates of IRS-NOMA are investigated in Section \ref{Ergodic Rate}. At last, the numerical results are presented to verify theoretical analyses in Section \ref{Numerical Results} and followed by concluding remarks in Section \ref{Conclusion}. The proofs of mathematics are collected in the Appendix.

The main notations used in this paper are shown as follows.
$\mathbb{E}\{\cdot\}$ denotes expectation operation. ${f_X}\left(  \cdot  \right)$ and ${F_X}\left(  \cdot  \right)$ denote the probability density function (PDF) and cumulative distribution function (CDF) of a random variable $X$, respectively. The superscripts ${\left(  \cdot  \right)^H}$ stand for the conjugate-transpose operation. diag$\left(  \cdot  \right)$ represents a diagonal matrix. $\otimes$ denotes the Kronecker product. ${\bf{I}}_P$ is a $P \times P$ identity matrix.

\section{Network Model}\label{System Model}
\subsection{Network Descriptions}
\begin{figure}[t!]
\centering
 \includegraphics[width= 3.48in, height=1.8in]{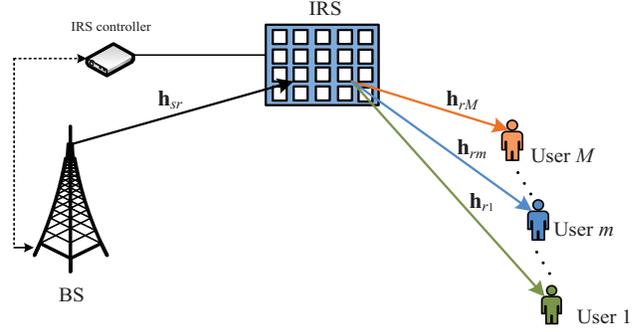}
 \caption{An IRS-assisted downlink NOMA network model, where the communications between the BS and terminal users are completed with the help of IRS.}
\label{System_Model_IRS_PLS_NOMA}
\end{figure}

Considering an IRS-assisted NOMA communication scenario as illustrated in Fig. \ref{System_Model_IRS_PLS_NOMA}, in which a base station (BS) sends the signals to $M$ terminal users with the assistance of an IRS. Assuming that the direct links between BS and users are assumed strongly attenuated and the communication can be only established through the IRS \cite{Emil2019,Huang8741198}. More specifically, the practical application of urban district scenarios can be supported by IRS-NOMA networks, where the communications between BS and users are blocked by the high-rise buildings. To provide the straightforward analyses, we assume that the BS and users are equipped single antenna, respectively\footnote{ It is noteworthy that multiple antennas equipped at the BS and user nodes will further enhance the performance of IRS-NOMA networks,  which is set aside for our future  treatise.}. The IRS is mounted with $K$ reconfigurable reflecting elements, which can be controlled by the communication oriented software. Additionally, the complex online control and signal processing operations are required for IRS-NOMA, but it is beyond the scope of this paper. The complex channel coefficient between the BS and IRS, and between the IRS and users are denoted as ${{\bf{h}}_{sr}} \in \mathbb{C}{^{K \times 1}}$ and ${{{\bf{h}}_{rm} }} \in \mathbb{C}{^{K \times 1}}$, respectively. In urban district scenarios, there are obstacles that can scatter a large number of radio signals. At this moment, the wireless links considered in IRS-NOMA networks are modeled as Rayleigh fading channels and our future work will relax this idealized simplifying assumption. These wireless links are perturbed by additive white Gaussian noise (AWGN) with the mean power $N_{0}$ simultaneously. Without loss of generality, the effective cascade channel gains from the BS to IRS and then to users are ordered as ${\left| {{\bf{h}}_{sr}^{H} { \bf{\Theta}} {{\bf{h}}_{r1}}} \right|^2} \le  \cdots  \le {\left| {{\bf{h}}_{sr}^{H}{ \bf{\Theta }} {{\bf{h}}_{rm}}} \right|^2} \le  \cdots  \le {\left| {{\bf{h}}_{sr}^{H}{ \bf{\Theta}} {{\bf{h}}_{rM}}} \right|^2}$ \cite{David2003Order,Yue8370069Unified}, where $ { \bf{\Theta }}  = {\rm{diag}}\left( {\beta {e^{j{\theta _1}}},...,\beta {e^{j{\theta _k}}},...,\beta {e^{j{\theta _K}}}} \right)$ is a diagonal matrix, $\beta  \in \left[ {0,1} \right]$ and ${\theta _k} \in \left[ {0,2\pi } \right)$ denote the fixed reflection amplitude coefficient and the phase shift of the $k$-th reflecting element of the IRS, respectively \cite{WuTowards2019,Ding2019IRS}.
We assume that the channel state information of all wireless channel are perfectly available at the BS. The imperfect channel state information is more suitable for evaluating practical scenarios, which will be set aside in our future work. Note that the IRS can be deployed at both low and high frequency bands, but it has greater advantages in the high frequency bands, which is conducive to coverage enhancement \cite{Kishk2020IRSdeployment,Choi2020IRS}. In this manuscript, the low frequency bands are taken into account for IRS-NOMA systems seriously.

\subsection{Signal Model}
The BS sends the superposed signals to $M$ users by the virtue of an IRS. Hence the received signal ${y_m}$ reflected by IRS at the $m$-th user is given by
\begin{align}\label{The received signal expression at the m-th user}
{y_m} = {\bf{h}}_{sr}^{H} { \bf{\Theta} } {{\bf{h}}_{rm}}\sum\limits_{i = 1}^M {\sqrt {{a_i}{P_s}} {x_i}}  + {n_m},
\end{align}
where $x_{i}$ is assumed to be normalised the unity power signal for the $i$-th user, i.e., $\mathbb{E}\{x_{i}^2\}= 1$. The $i$-th user's power allocation factor $a_i$ satisfies the relationship ${a_1} \ge \cdots \ge {a_{m}} \ge \cdots \ge {a_{M}}$ with $\mathop \sum \limits_{i = 1}^M {a _i} = 1$, which is for the sake of user fairness. The optimal design of power allocation coefficients between users will further heighten the performance of NOMA networks, but it is beyond the scope of this paper. $P_{s}$ denotes the normalized transmission power at the BS.
${{\bf{h}}_{sr}} = {\left[ {h_{sr}^1 \cdots h_{sr}^k \cdots h_{sr}^K} \right]^H}$, where $h_{sr}^k \sim {\cal C}{\cal N}\left( {0,{\Omega _{sr}}} \right)$  denotes the complex channel coefficient from the BS to the $k$-th reflecting element of IRS. ${{\bf{h}}_{rm}} = {\left[ {h_{rm}^1 \cdots h_{rm}^k \cdots h_{rm}^K} \right]^H}$, where $h_{rm}^k  \sim {\cal C}{\cal N}\left( {0,{\Omega _{rm}}} \right)$ denotes the complex channel coefficient from the $k$-th reflecting element of IRS to the $m$-th user. ${n_m}$ denotes AWGN at the $m$-th user.

On the basis of NOMA principle, the received signal-to-interference-plus-noise ratio (SINR) at the $m$-th user to detect the $q$-th user's information $( m \ge q)$ is given by
\begin{align}\label{The SINR of the m-th user to detect the p-th user}
{\gamma _{m \to q}} =  \frac{{ \rho {{\left| {{\bf{h}}_{sr}^{H} { \bf{\Theta}} {{\bf{h}}_{rm}}} \right|}^2}{a_q} }}{{ \rho {{\left| {{\bf{h}}_{sr}^{H} { \bf{\Theta}} {{\bf{h}}_{rm}}} \right|}^2} \sum\limits_{i = q + 1}^M {{a_i} + \varpi \rho {{\left| {{h_I}} \right|}^2}} + 1}},
\end{align}
where ${\rho}=\frac{{{P_s}}}{{ N_{0}}}$ denotes the transmit SNR and $\varpi  \in \left[ {0,1} \right]$. More precisely, $\varpi {\rm{ = 0}}$ and $\varpi {\rm{ = 1}}$  denote the pSIC and ipSIC operations. Without loss of generality,  assuming that the residual interference from ipSIC is modeled as the Rayleigh fading and corresponding complex channel coefficient is denoted by ${{h_I}}  \sim {\cal C}{\cal N}\left( {0,{\Omega _I}} \right)$.

After striking out the previous $M-1$ users' signals with SIC, the received SINR at the $M$-th user to detect its own information can be given by
\begin{align}\label{The SINR of the M-th user}
{\gamma _M} = \frac{{\rho {{\left| {{\bf{h}}_{sr}^{H} { \bf{\Theta}} {{\bf{h}}_{rM}}} \right|}^2}{a_M} }}{{\varpi \rho {{\left| {{h_I}} \right|}^2} + 1}}.
\end{align}

\subsection{IRS-NOMA with 1-bit Coding}
From the perspective of practical communication applications, continuously changing the reflection amplitude and phase shift of each IRS's element is beneficial to enhance the network performance. This alternative implement needs the accurate design and expensive hardware architecture, which will result in the higher cost of IRS. To facilitate implement and analysis, 1-bit coding scheme is selected to achieve the discrete amplitude/phase shift levels for IRS-assisted NOMA networks \cite{Cui2014,Ding2019IRS}, where the elements of diagonal matrix $\bf{\Theta}$ are replaced with 0 or 1. It is the scalable and cost-effective solution with the number of reflecting elements becomes larger.

Assuming that the number reflecting elements $K$ of IRS are equal to $PQ$, i.e., $K=PQ$, where $P$ and $Q$ are integers. Define $\bf{V} = {{\bf{I}}_P} \otimes {{\bf{1}}_Q}$, where ${{\bf{1}}_Q}$ is a column vector of all ones with $Q \times 1$.  The $p$-th column of $\bf{V}$ is denoted by ${{{\bf{v}}_p}}$ with $K \times 1$ and ${\bf{v}}_p^H {{\bf{v}}_l}= 0$ for $p \ne l$. As a result, the SINRs of \eqref{The SINR of the m-th user to detect the p-th user} and \eqref{The SINR of the M-th user} can be maximized by randomly choosing one column i.e., ${{{\bf{v}}_p}}$ from $\bf{V}$ and rewritten as
\begin{align}\label{maximize the SINR the m-th user}
{{\tilde \gamma }_{m \to q}} = \mathop {\max }\limits_{{{\bf{v}}_p}} \frac{{\rho {{\left| {{\bf{v}}_p^H{{\bf{D}}_m}{{\bf{h}}_{sr}}} \right|}^2}{a_q}}}{{\rho {{\left| {{\bf{v}}_p^H{{\bf{D}}_m}{{\bf{h}}_{sr}}} \right|}^2}\sum\limits_{i = q + 1}^M {{a_i}}  + {\varpi \rho {{\left| {{h_I}} \right|}^2}} + 1}},
\end{align}
and
\begin{align}\label{maximize the SINR the M-th user}
{{\tilde \gamma }_M} = \mathop  {\max }\limits_{{{\bf{v}}_p}} \frac{{\rho {{\left| {{\bf{v}}_p^H{{\bf{D}}_M}{{\bf{h}}_{sr}}} \right|}^2}{a_M}}}{{\varpi \rho {{\left| {{h_I}} \right|}^2} + 1}},
\end{align}
respectively, where ${{\bf{D}}_m}$ and ${{\bf{D}}_M}$ are the diagonal matrix with its diagonal elements obtained from  ${{\bf{h}}_{rm}}$ and ${{\bf{h}}_{rM}}$.  The following network performance of IRS-NOMA is discussed with 1-bit coding scheme.
\subsection{IRS-OMA}
In this subsection, the IRS-OMA scheme is regarded as one of the benchmarks for comparison purpose, where an IRS is deployed to assist in the transmission from the BS to a user $d$. On the condition of the above assumptions, the maximized SNR of user $d$ with 1-bit coding scheme for IRS-OMA can be given by
\begin{align}\label{maximize the SINR the d-th user}
{\gamma _d} = \mathop {\max }\limits_{{{\bf{v}}_p}} \rho {\left| {{\bf{v}}_p^H{{\bf{D}}_d}{{\bf{h}}_{sr}}} \right|^2},
\end{align}
where ${{\bf{h}}_{rd}} = {\left[ {h_{rd}^1 \cdots h_{rd}^k \cdots h_{rd}^K} \right]^H}$, $h_{rd}^k  \sim {\cal C}{\cal N}\left( {0,{\Omega _{rd}}} \right)$ denotes the complex channel coefficient from the $k$-th reflecting element of IRS to user $d$. ${{\bf{D}}_d}$ and is the diagonal matrix with its diagonal elements obtained from  ${{\bf{h}}_{rd}}$.

\section{Outage probability}\label{Outage probability}
As mentioned in conventional NOMA, the SIC scheme is carried out at the $m$-th user  by decoding and striking out
the $q$-th user's information $\left( {m \ge  q \ge 1} \right)$ before it detects its own signal. If the $m$-th user cannot successfully detect the $q$-th user's information, an outage occurs and is denoted by
\begin{align}\label{Events}
{{\rm{E}}_{m,q}} = \left\{ {\frac{{\rho {{\left| {{\bf{v}}_p^H{{\bf{D}}_m}{{\bf{h}}_{sr}}} \right|}^2}{a_q}}}{{\rho {{\left| {{\bf{v}}_p^H{{\bf{D}}_m}{{\bf{h}}_{sr}}} \right|}^2}\sum\limits_{i = q + 1}^M {{a_i}}  + \varpi \rho {{\left| {{h_I}} \right|}^2} + 1}} < {\gamma _{t{h_q}}}} \right\},
\end{align}
where ${\gamma _{t{h_q}}} = {2^{{R_q}}} - 1$ with ${{{R_q}}}$ being the target rate at the $m$-th user to detect $x_q$.  Note that the first user i.e., $m=1$ does not carry out the SIC procedure and there is no residual interference term in the above equation.
As a consequence, the outage probability of $m$-th user with 1-bit coding for IRS-NOMA networks can be expressed as
\begin{align}\label{the OP expression}
{P_m} = {\rm{Pr}}\left[ {{\rm{min}}\left( {{{\rm{E}}_{m,1}},{{\rm{E}}_{m,2}}, \cdots ,{{\rm{E}}_{m,m}}} \right)} \right].
\end{align}
It is worth pointing out that the first user (i.e., $m=1$) with the worse channel condition does not execute the SIC procedure.

\begin{theorem}\label{Theorem1:the OP of the m-th user with ipSIC under Rayleigh fading channel}
Under Rayleigh fading channels, the closed-form expression for outage probability of the $m$-th user with ipSIC in IRS-NOMA networks is given by
\begin{align}\label{the OP of the m-th user with ipSIC under Rayleigh fading channel}
P_{ipSIC}^{m} \approx & \left\{ {{\phi _m}\sum\limits_{l = 0}^{M - m} {\sum\limits_{u = 1}^U {{
   {M - m}  \choose
   l  }} } \frac{{{{\left( { - 1} \right)}^l}{H_u}}}{{m + l}}} \right.\left[ {1 - \frac{2}{{\Gamma \left( Q \right)}}} \right.  \nonumber \\
& {\left. {{{\left. { \times {{\left( {\frac{{\psi _m^ * \Lambda }}{{{\Omega _{sr}}{\Omega _{rm}}}}} \right)}^{\frac{Q}{2}}}{K_Q}\left( {2\sqrt {\frac{{\psi _m^ * \Lambda }}{{{\Omega _{sr}}{\Omega _{rm}}}}} } \right)} \right]}^{m + l}}} \right\}^P} ,
\end{align}
where ${\phi _m} = \frac{{M!}}{{\left( {M - m} \right)!\left( {m - 1} \right)!}}$,  $\psi _m^ *  = \max \left\{ {{\psi _1}, \ldots ,{\psi _m}} \right\}$ and ${\psi _m} = \frac{{{\gamma _{t{h_m}}}}}{{\rho \left( {{a_m} - {\gamma _{t{h_m}}}\sum\nolimits_{i = m+ 1}^M {{a_i}} } \right)}}$ with ${a_m} > {\gamma _{t{h_m}}}\sum\nolimits_{i = m + 1}^M {{a_i}} $, ${\psi _M} = \frac{{{\gamma _{t{h_M}}}}}{{\rho {a_M}}}$, ${\gamma _{t{h_m}}} = {2^{{R_m}}} - 1$ with ${{{R_m}}}$ being the target rate at the $m$-th user to detect $x_m$. 
 $\Lambda  = \left( {\varpi \rho {\Omega _I}{x_u} + 1} \right)$. ${H_u}$ and ${{r_u}}$ are the weight and abscissas for Gauss-Laguerre integration, respectively. More specifically, ${{{ r}_u}}$ is the $w$-th zero of Laguerre polynomial ${{ L}_U}\left( {{{ r}_u}} \right)$ and the corresponding $w$-th weight is given by ${H_u} = \frac{{{{\left( {U!} \right)}^2}{r_u}}}{{{{\left[ {{L_{U + 1}}\left( {{r_u}} \right)} \right]}^2}}}$. The parameter $U$ is to ensure a complexity-accuracy tradeoff. ${K_v}\left(  \cdot  \right)$ is the modified Bessel function of the second kind with order $v$. $\Gamma \left(  \cdot  \right)$ denotes the gamma function \cite[Eq. (8.310.1)]{2000gradshteyn}.
\begin{proof}
See Appendix~A.
\end{proof}
\end{theorem}

\begin{corollary}\label{Corollary:the OP of the m-th user with ipSIC under Rayleigh fading channel}
For the special case with substituting $\varpi=0$ into \eqref{AppendixA: OP expression of the m-th user with Rice fading}, the closed-form expression for outage probability of the $m$-th user with pSIC in IRS-NOMA networks is given by
\begin{align}\label{the OP of the m-th user with pSIC under Rayleigh fading channel}
P_{pSIC}^{m} =& \left\{ {{\phi _m}\sum\limits_{l = 0}^{M - m} {{
   {M - m}  \choose
   l }} \frac{{{{\left( { - 1} \right)}^l}}}{{m + l}}\left[ {1 - \frac{2}{{\Gamma \left( Q \right)}}} \right.} \right. \nonumber\\
 &{\left. {{{\left. { \times {{\left( {\frac{{\psi _m^ * }}{{{\Omega _{sr}}{\Omega _{rm}}}}} \right)}^{\frac{Q}{2}}}{K_Q}\left( {2\sqrt {\frac{{\psi _m^ * }}{{{\Omega _{sr}}{\Omega _{rm}}}}} } \right)} \right]}^{m + l}}} \right\}^P} .
\end{align}
\end{corollary}

For IRS-OMA, an outage is defined as the probability that the instantaneous SNR $({\gamma _d})$ falls bellow a threshold SNR ${\gamma _{t{h_d}}}$. Hence the outage probability of user $d$ with 1-bit coding which can be expressed as
\begin{align}\label{outage expression for OMA}
{P_d} = {\rm{Pr}}\left( {{\gamma _d} \le {\gamma _{t{h_d}}}} \right),
\end{align}
where ${\gamma _{t{h_d}}} = {2^{{R_{oma}}}} - 1$ with ${R_{oma}}$ being the target rate of user $d$ to detect $x_d$. Referring to \eqref{the OP of the m-th user with pSIC under Rayleigh fading channel} and removing the order operation, we can derive the outage probability for IRS-OMA in the following corollary.
\begin{corollary}\label{Corollary:the OP of OMA}
The closed-form expression of outage probability for IRS-OMA networks with 1-bit coding is given by
\begin{align}\label{outage of OMA}
{P_d} = {\left[ {1 - \frac{2}{{\Gamma \left( Q \right)}}{{\left( {\frac{{{\gamma _{t{h_d}}}}}{{\rho {\Omega _{sr}}{\Omega _{rd}}}}} \right)}^{\frac{Q}{2}}}{K_Q}\left( {2\sqrt {\frac{{{\gamma _{t{h_d}}}}}{{\rho {\Omega _{sr}}{\Omega _{rd}}}}} } \right)} \right]^P}.
\end{align}
\end{corollary}
\subsection{Diversity Analysis}\label{Diversity Analysis}
In order to gain better insights, the diversity order is usually selected to evaluate the outage behaviors for communication systems, which is able to describe how fast the outage probability decreases with the transmitting SNR. Hence the diversity order can be expressed as
\begin{align}\label{The definition of diversity order for IRS-NOMA}
d =  - \mathop {\lim }\limits_{\rho  \to \infty } \frac{{\log \left( {P_\infty \left( \rho  \right)} \right)}}{{\log \rho }},
\end{align}
where ${P_\infty \left( \rho  \right)}$ denotes the asymptotic outage probability in the high SNR regime.

\begin{corollary}\label{corollary:the OP of the m-th user with ipSIC under Rayleigh fading channel}
Based on analytical result in \eqref{the OP of the m-th user with ipSIC under Rayleigh fading channel}, when $\rho  \to \infty$, the asymptotic outage probability of the $m$-th user with ipSIC for IRS-NOMA networks is given by
\begin{align}\label{the diversity order of the m-th user with ipSIC under Rayleigh fading channel}
 P_{ipSIC }^{m,\infty} \approx & \left\{ {\frac{{M!}}{{\left( {M - m} \right)!m!}}\sum\limits_{w = 1}^W {{H_w}\left[ {1 - \frac{2}{{\Gamma \left( Q \right)}}} \right.} } \right. \nonumber \\
 &{\left. {{{\left. { \times {{\left( {\frac{{\varpi \vartheta _m^ * {\Omega _I}{x_w}}}{{{\Omega _{sr}}{\Omega _{rm}}}}} \right)}^{\frac{Q}{2}}}{K_Q}\left( {2\sqrt {\frac{{\varpi \vartheta _m^ * {\Omega _I}{x_w}}}{{{\Omega _{sr}}{\Omega _{rm}}}}} } \right)} \right]}^m}} \right\}^P} ,
\end{align}
where $\vartheta _m^ *  = \max \left\{ {{\vartheta _1}, \ldots ,{\vartheta _m}} \right\}$ and ${\vartheta _m} = \frac{{{\gamma _{t{h_m}}}}}{{\left( {{a_m} - {\gamma _{t{h_m}}}\sum\nolimits_{i = m + 1}^M {{a_i}} } \right)}}$ with ${a_m} > {\gamma _{t{h_m}}}\sum\nolimits_{i = m + 1}^M {{a_i}} $.
\end{corollary}
\begin{remark}\label{Remark1:the diversity order of the m-th user with ipSIC under Rayleigh fading channel}
Upon substituting \eqref{the diversity order of the m-th user with ipSIC under Rayleigh fading channel} into \eqref{The definition of diversity order for IRS-NOMA}, the diversity order of the $m$-th user with ipSIC for IRS-NOMA is equal to $zero$. This is due to the influence of residual interference from ipSIC.
\end{remark}

\begin{corollary}\label{corollary:the OP of the m-th user with pSIC under Rayleigh fading channel for Q=1}
For the cases ${Q =1}$ and ${Q \ge 2}$,  the asymptotic outage probability of the $m$-th user with pSIC at high SNRs are given by
\begin{align}\label{the diversity order of the m-th user with pSIC under Rayleigh fading channel for Q1}
P_{pSIC }^{m,\infty} =& \left\{ {\frac{{M!}}{{\left( {M - m} \right)!m!}}\left[ { - \frac{{2\psi _m^ * }}{{{\Omega _{sr}}{\Omega _{rm}}}}} \right.} \right. \nonumber \\
 &{\left. {{{\left. { \times \ln \left( {\sqrt {\frac{{\psi _m^ * }}{{{\Omega _{sr}}{\Omega _{rm}}}}} } \right)} \right]}^m}} \right\}^K} ,Q = 1  ,
\end{align}
and
\begin{align}\label{the diversity order of the m-th user with pSIC under Rayleigh fading channel}
P_{pSIC }^{m,\infty} =& \left\{ {\frac{{M!}}{{\left( {M - m} \right)!m!}}} \right. \nonumber \\
 &{\left. { \times {{\left( {\frac{{\psi _m^ * }}{{\left( {Q - 1} \right){\Omega _{sr}}{\Omega _{rm}}}}} \right)}^m}} \right\}^P},Q \ge 2 ,
\end{align}
respectively.
\begin{proof}
To facilitate the calculation, we employ the series representation of Bessel functions  ${K_v}\left(  x  \right)$ to obtain the high SNR approximation. When $v=1$ and $v \ge 2$,  ${K_v}\left(  x  \right)$ can be approximated as
\begin{align}\label{K_1}
{K_1}\left( x \right) \approx \frac{1}{x} + \frac{x}{2}\ln \left( {\frac{x}{2}} \right),
\end{align}
and
\begin{align}\label{K_Q}
{K_v}\left( x \right) \approx \frac{1}{2}\left[ {\frac{{{2^v}\left( {v - 1} \right)!}}{{{x^v}}} - \frac{{{2^{v - 2}}\left( {v - 2} \right)!}}{{{x^{v - 2}}}}} \right],
\end{align}
respectively.
Upon substituting \eqref{K_1} and \eqref{K_Q} into \eqref{the OP of the m-th user with pSIC under Rayleigh fading channel}, and then taking the first term $(l=0)$ of summation term, we can obtain \eqref{the diversity order of the m-th user with pSIC under Rayleigh fading channel for Q1} and \eqref{the diversity order of the m-th user with pSIC under Rayleigh fading channel}, respectively. The proof is completed.
\end{proof}
\end{corollary}
\begin{remark}\label{Remark2:the diversity order of the m-th user with pSIC under Rayleigh fading channel Q1}
Upon substituting \eqref{the diversity order of the m-th user with pSIC under Rayleigh fading channel for Q1} and \eqref{the diversity order of the m-th user with pSIC under Rayleigh fading channel} into \eqref{The definition of diversity order for IRS-NOMA}, the diversity orders of the $m$-th user with pSIC for cases $Q=1$ and ${Q \ge 2}$ are $mK$ and $mP$, respectively. As can be observed that the diversity order of the $m$-th user are in connection with the number of reflecting elements of IRS and channel ordering.
\end{remark}

\begin{corollary}\label{corollary:the diversity of OP for OMA}
Similar to the procedures in \eqref{the diversity order of the m-th user with pSIC under Rayleigh fading channel for Q1} and \eqref{the diversity order of the m-th user with pSIC under Rayleigh fading channel}, the asymptotic outage probability of IRS-OMA for both ${Q =1}$ and ${Q \ge 2}$ at high SNRs are given by
\begin{align}\label{the diversity order of OMA for Q1}
P_d^\infty  = {\left[ { - 2\left( {\frac{{{\gamma _{t{h_d}}}}}{{\rho {\Omega _{sr}}{\Omega _{rd}}}}} \right)\ln \left( {\sqrt {\frac{{{\gamma _{t{h_d}}}}}{{\rho {\Omega _{sr}}{\Omega _{rd}}}}} } \right)} \right]^K},Q = 1,
\end{align}
and
\begin{align}\label{the diversity order of OMA for Q2}
P_d^\infty  = {\left[ {\frac{{{\gamma _{t{h_d}}}}}{{\rho \left( {Q - 1} \right){\Omega _{sr}}{\Omega _{rd}}}}} \right]^P},Q \ge 2,
\end{align}
respectively.
\end{corollary}
\begin{remark}\label{Remark2:the diversity order of OMA}
Upon substituting \eqref{the diversity order of OMA for Q1} and \eqref{the diversity order of OMA for Q2} into \eqref{The definition of diversity order for IRS-NOMA}, the diversity orders of IRS-OMA with 1-bit coding for cases $Q=1$ and ${Q \ge 2}$ are $K$ and $P$, respectively.
\end{remark}
\subsection{Delay-Limited Transmission}\label{delay-limited mode System throughput}
In the delay-limited transmission mode, the BS sends the information at a constant rate, which is subject to outage according to the random fading of wireless channels \cite{Nasir6552840,Zhong7572025}. Hence the system throughput of IRS-NOMA with ipSIC/pSIC in the delay-limited transmission mode can be given by
\begin{align}\label{the system throughput of delay-limited mode}
{R_{m,dl}} = \sum\limits_{m = 1}^M {\left( {1 - P_\xi ^m} \right){R_m}} ,
\end{align}
where $\xi  \in \left\{ {ipSIC,pSIC} \right\}$. $P_{m}^{ipSIC}$ and $P_{m}^{pSIC}$ can be obtain from \eqref{the OP of the m-th user with ipSIC under Rayleigh fading channel} and \eqref{the OP of the m-th user with pSIC under Rayleigh fading channel}, respectively.
\section{Ergodic Rate}\label{Ergodic Rate}
In this section, the ergodic rate of the $m$-th user with ipSIC/pSIC for IRS-NOMA networks is discussed in detail, where the target rates of users are determined by the channel conditions. The $m$-th user detects the $p$-th user's information successfully, since it holds ${\left| {{\bf{h}}_{sr}^H\Theta {{\bf{h}}_{rm}}} \right|^2} \ge {\left| {{\bf{h}}_{sr}^H\Theta {{\bf{h}}_{rp}}} \right|^2}$. Under this situation, the achievable rate of the $m$-th user can be written as ${{\tilde R}_m} = \log \left( {1 + {{\tilde \gamma }_{m \to m}}} \right)$.
Based on \eqref{maximize the SINR the m-th user} and \eqref{maximize the SINR the M-th user}, the ergodic rates of the $m$-th user and $M$-th user with ipSIC for IRS-NOMA networks can be given by
\begin{align}\label{the ergodic rate of the m-th user ipSIC}
&R_{m,erg}^{ipSIC} =  \nonumber \\
&{\mathbb{E}}\left\{ {\log \left( {1 + \mathop {\max }\limits_{{{\bf{v}}_p}} \frac{{\rho {{\left| {{\bf{v}}_p^H{{\bf{D}}_m}{{\bf{h}}_{sr}}} \right|}^2}{a_m}}}{{\rho {{\left| {{\bf{v}}_p^H{{\bf{D}}_m}{{\bf{h}}_{sr}}} \right|}^2}{{\bar a}_m} + \varpi \rho {{\left| {{h_I}} \right|}^2} + 1}}} \right)} \right\},
\end{align}
and
\begin{align}\label{the ergodic rate of the M-th user ipSIC}
R_{M,erg}^{ipSIC} = {\mathbb{E}}\left\{ {\log \left( {1 + \mathop {\max }\limits_{{{\bf{v}}_p}} \frac{{\rho {{\left| {{\bf{v}}_p^H{{\bf{D}}_M}{{\bf{h}}_{sr}}} \right|}^2}{a_M}}}{{\varpi \rho {{\left| {{h_I}} \right|}^2} + 1}}} \right)} \right\},
\end{align}
respectively, where $\varpi  = 1$ and ${{\bar a}_m} = \sum\nolimits_{i = m + 1}^M {{a_i}} $. One can be seen from the equations, there are no the closed-form solutions. However, these expressions can be evaluated numerically by using the standard softwares such as Matlab or Mathematica. By employing pSIC, the ergodic rates of the $m$-th user and $M$-th user are presented in the following part.

\begin{theorem}\label{Theorem:Ergodic Rate of the m-th user}
For the special case with substituting $\varpi  = 0$ into \eqref{the ergodic rate of the m-th user ipSIC}, the closed-form expression of ergodic rate for the $m$-th user with pSIC in IRS-NOMA networks is given by
\begin{align}\label{Ergodic Rate of the m-th user}
&R_{m,erg}^{pSIC} \approx \frac{{\pi {a_m}}}{{N\ln 2}}\sum\limits_{n = 1}^N {\frac{{\sqrt {1 - {x_n}^2} }}{{2{{\bar a}_m} + \left( {1 + {x_n}} \right){a_m}}}\left\langle {1 - \left\{ {{\phi _m}} \right.} \right.} \nonumber \\
&  \times \sum\limits_{l = 0}^{M - m} {\sum\limits_{r = 0}^{m + l} {{
   {M - m}  \choose
   l  }} {
   {m + l}  \choose
   r  }\frac{{{{\left( { - 1} \right)}^{l{\rm{ + }}r}}}}{{m + l}}{{\left( {\frac{2}{{\Gamma \left( Q \right)}}} \right)}^r}}  \nonumber\\
&  \times \left. {{{\left. {{{\left[ {{{\left( {\frac{{{\varphi _m}\left( {1 + {x_n}} \right)}}{{{{\bar a}_m}\left( {1 - {x_n}} \right)}}} \right)}^{\frac{Q}{2}}}{K_Q}\left( {2\sqrt {\frac{{{\varphi _m}\left( {1 + {x_n}} \right)}}{{{{\bar a}_m}\left( {1 - {x_n}} \right)}}} } \right)} \right]}^r}} \right\}}^P}} \right\rangle ,
\end{align}
where ${\phi _m} = \frac{{M!}}{{\left( {M - m} \right)!\left( {m - 1} \right)!}}$, ${\varphi _m} = \frac{1}{{\rho {\Omega _{sr}}{\Omega _{rm}}}}$, ${x_n} = \cos \left( {\frac{{2n - 1}}{{2N}}\pi } \right)$ and $N$ is a parameter to ensure a complexity-accuracy tradeoff.
\begin{proof}
See Appendix~B.
\end{proof}
\end{theorem}

\begin{theorem}\label{Theorem:Ergodic Rate of the M-th user}
For the special case with substituting $\varpi  = 0$ into \eqref{the ergodic rate of the M-th user ipSIC}, the exact expression of ergodic rate for the $M$-th user with pSIC in IRS-NOMA networks is given by
\begin{align}\label{Ergodic Rate of the M-th user}
R_{M,erg}^{pSIC} =& \frac{{\rho {a_M}}}{{\ln 2}}\sum\limits_{r = 1}^{MP} {{
   {MP}  \choose
   r  }\frac{{{{\left( { - 1} \right)}^{r + 1}}{2^r}}}{{{{\left[ {\Gamma \left( Q \right){{\left( {{\Omega _{sr}}{\Omega _{rm}}} \right)}^{\frac{Q}{2}}}} \right]}^r}}}} \nonumber \\
&  \times \int_0^\infty  {\frac{{{x^{\frac{{rQ}}{2}}}{{\left[ {{K_Q}\left( {2\sqrt {\frac{x}{{{\Omega _{sr}}{\Omega _{rm}}}}} } \right)} \right]}^r}}}{{1 + \rho {a_M}x}}dx}  .
\end{align}
\begin{proof}
See Appendix~C.
\end{proof}
\end{theorem}

For IRS-OMA, based on \eqref{maximize the SINR the d-th user}, the ergodic rate of user $d$ with 1-bit coding can be expressed as
\begin{align}\label{the ergodic rate for user d OMA}
{R_{d,erg}} = {\mathbb{E}}\left[ {\log \left( {1 + \mathop {\max }\limits_{{{\bf{v}}_p}} \rho {{\left| {{\bf{v}}_p^H{{\bf{D}}_d}{{\bf{h}}_{sd}}} \right|}^2}} \right)} \right].
\end{align}
\begin{corollary}
Similar to the derivation process in \eqref{Ergodic Rate of the M-th user}, the exact expression of ergodic rate for IRS-OMA with 1-bit coding is given by
\begin{align}\label{the ergodic rate for OMA}
 {R_{d,erg}} = &\frac{\rho }{{\ln 2}}\sum\limits_{r = 1}^P {{
   P  \choose
   r  }\frac{{{{\left( { - 1} \right)}^{r + 1}}{2^r}}}{{{{\left[ {\Gamma \left( Q \right){{\left( {{\Omega _{sr}}{\Omega _{rd}}} \right)}^{\frac{Q}{2}}}} \right]}^r}}}} \nonumber  \\
  & \times \int_0^\infty  {\frac{{{x^{\frac{{rQ}}{2}}}{{\left[ {{K_Q}\left( {2\sqrt {\frac{x}{{{\Omega _{sr}}{\Omega _{rd}}}}} } \right)} \right]}^r}}}{{1 + \rho x}}dx}  .
\end{align}
\end{corollary}
\subsection{Slope Analysis}
Similar to the diversity order, the high SNR slope aims to capture the diversification of ergodic rate with the transmitting SNRs, which can be defined as
\begin{align}\label{high SNR slope}
S = \mathop {\lim }\limits_{\rho  \to \infty } \frac{{R_{m}^\infty \left( \rho  \right)}}{{\log \left( \rho  \right)}},
\end{align}
where ${{R_{m}^\infty \left( \rho  \right)}}$ denotes the asymptotic ergodic rate in the high SNR regime.

According to \eqref{AppendixC:The ergodic rate expression with pSIC}, when $\rho  \to \infty $, the asymptotic ergodic rate of the $m$-th user with pSIC is given by
\begin{align}\label{asymptotic ergodic rate of the $m$-th user}
R_{erg}^{m,\infty } = \log \left[ {1 + \left( {\frac{{{a_m}}}{{{{\bar a}_m}}}} \right)} \right].
\end{align}
\begin{remark}\label{Remark3}
Upon substituting \eqref{asymptotic ergodic rate of the $m$-th user} into \eqref{high SNR slop}, the high SNR slope of the $m$-th user with pSIC for IRS-NOMA networks is $zero$, which is the same as conventional NOMA.
\end{remark}

As can be seen from \eqref{Ergodic Rate of the M-th user} that the exact derivation of the approximation at high SNRs appears mathematically intractable. Furthermore, we focus our attention on evaluating the slope of ergodic rate for the $M$-th user via the assistance of its upper bound. By noticing that $\log \left( {1 + {x^2}} \right)$ is a concave function for $x \ge 0$, we invoke the Jensen’s inequality to derive an upper bound as
\begin{align}\label{upper bound of ergodic rate for the M-th user}
R_{M,erg}^{pSIC} \le \log \left[ {1 + \rho {a_M}{\mathbb{E}}\left( {\mathop {\max }\limits_{{{\bf{v}}_p}} {{\left| {{\bf{v}}_p^H{{\bf{D}}_M}{{\bf{h}}_{sr}}} \right|}^2}} \right)} \right].
\end{align}
We performe the derivative operation on \eqref{AppendixC:The CDF for M-th user with pSIC} in Appendix~C and some manipulates, the upper bound $R_{M,erg}^{pSIC,ub}$ is given by
\begin{align}\label{the final upper bound of ergodic rate for the M-th user}
R_{M,erg}^{pSIC,ub} = \log \left\{ {1 + \frac{{2MP\Phi {a_M}\rho }}{{\Gamma \left( Q \right){{\left( {\sqrt {{\Omega _{sr}}{\Omega _{rM}}} } \right)}^{Q + 1}}}}} \right\},
\end{align}
where $\Phi  = \int_0^\infty  {{{\left[ {1 - \frac{2}{{\Gamma \left( Q \right)}}{{\left( {\frac{x}{{{\Omega _{sr}}{\Omega _{rM}}}}} \right)}^{\frac{Q}{2}}}{K_Q}\left( {\sqrt {\frac{{4x}}{{{\Omega _{sr}}{\Omega _{rM}}}}} } \right)} \right]}^{MP - 1}}}   \\
\times {x^{\frac{{Q - 1}}{2}}}{K_{Q - 1}}\left( {2\sqrt {\frac{x}{{{\Omega _{sr}}{\Omega _{rM}}}}} } \right)dx  $ is constant with increasing the SNRs. Upon substituting \eqref{the final upper bound of ergodic rate for the M-th user} into \eqref{high SNR slop} and further applying  L'hospital rule, we can obtain the high SNR slope of ergodic rate for the $M$-th user in the following remark.
\begin{remark}\label{Remark4}
Upon substituting \eqref{the final upper bound of ergodic rate for the M-th user} into \eqref{high SNR slop}, the high SNR slope of the $M$-th user with pSIC for IRS-NOMA networks is equal to $one$. One can observe that the use of IRS to NOMA donot improve the slope of ergodic rate for the $M$-th user.
\end{remark}

\subsection{Delay-Tolerate Transmission}\label{delay-tolerate mode System throughput}
In the delay-tolerant mode, the BS sends the information at any fixed rate upper bounded by the ergodic capacity. Hence the system throughput of IRS-NOMA with pSIC is given by
\begin{align}\label{the system throughput of delay-tolerate  mode}
{R_{\zeta ,dt}} = \sum\limits_{m = 1}^M {R_{m,erg}^{pSIC}} ,
\end{align}
where $R_{m,erg}^{pSIC}$ and $R_{M,erg}^{pSIC}$ can be obtained from \eqref{Ergodic Rate of the m-th user} and \eqref{Ergodic Rate of the M-th user}, respectively.

To facilitate comparison, the diversity orders and high SNR slopes of the $m$-th user with ipSIC/pSIC for IRS-NOMA are summarized in TABLE I, where we use ``D'' and ``S'' to represent the diversity order and high SNR slope, respectively.
\begin{table}[!h]
\begin{center}
{\tabcolsep10pt\begin{tabular}{|l|l|l|l|l|}\hline   
  \textbf{Mode} & \textbf{SIC}  &\textbf{User} & \textbf{D} & \textbf{S} \\
     \hline
\multirow{2}{*}{IRS-OMA}    & \multirow{2}{*}{------}   & User $d$ $(Q=1)$ & $ K$ & $  - $ \\
\cline{3-5}
                     &  & User $d$ $(Q \ge 2)$  & $ P$ & $ - $  \\
\hline
\multirow{4}{*}{IRS-NOMA}  & \multirow{1}{*}{ipSIC} & User $m$& $ 0$ & $ - $ \\
\cline{2-5}
               &\multirow{2}{*}{pSIC}  &User $m$ $(Q=1)$ & $ mK$   & $0$  \\
\cline{3-5}
               &  &User $m$ $(Q \ge 2)$  & $ mP$   & $  0$ \\
\hline

\end{tabular}}{}
\label{tab1}
\end{center}
\caption{Diversity order and high SNR slope for IRS-NOMA networks.}
\end{table}

\section{Energy Efficiency}\label{Energy Efficiency}
In IRS-NOMA networks, the total power consumption is composed of the BS transmit power, the hardware static power dissipated at the BS, IRS and user terminals \cite{Huang8741198}, which can be written as
\begin{align}
{P_{total}} = \kappa {P_S} + {P_{BS}} + {P_{IRS}} + \sum\limits_{m = 1}^M {{P_{UE,m}}},
\end{align}
where $\kappa  = {\nu ^{ - 1}}$ with $\nu$ being the efficiency of the transmitting power amplifier, ${P_{S}}$ denotes the transmitting power of the BS. $P_{BS}$ and $ {P_{IRS}} = K{P_k}\left( b \right) $ represent the total hardware static dissipated power at the BS and IRS, respectively, where ${P_k}\left( b \right) $ is the power consumption of each phase shifter having $b$-bit resolution \cite{Rusu2016shifters,Ribeiro2018MIMO}. ${{P_{m,UE}}}$ denotes the hardware power dissipated of the $m$-th user.   As a further development,  the energy efficiency (EE) of IRS-NOMA can be interpreted as the sum data rate divided by the total power consumption i.e., ${\eta_{EE} } = \frac{{{\rm{Sum~data~rate}}}}{{{\rm{Total~power~consumption}}}}$, and can be given by
\begin{align}\label{The Energy Efficiency of IRS-NOMA}
{\eta _{EE}} = \frac{{{R_\Phi }}}{{{P_{total}}}},
\end{align}
where ${R_\Phi }  \in  \left( {R_{m,dl}} , {R_{\zeta,dt}}\right)$.  ${R_{m,dl}}$ and ${R_{\zeta,dt}}$ can be obtained from \eqref{the system throughput of delay-limited mode} and \eqref{the system throughput of delay-tolerate  mode}, respectively.

\begin{table}[!t]
\centering
\caption{Table of Parameters for Numerical Results}
\tabcolsep5pt
\renewcommand\arraystretch{1.1} 
\begin{tabular}{|l|l|}
\hline
Monte Carlo simulations repeated  &  ${10^6}$ iterations \\
\hline
Pass loss exponent  & $\alpha=2$  \\
\hline
\multirow{3}{*}{The power allocation factors for users }&  \multirow{1}{*}{$a_1=0.5$}   \\
                                                        &  \multirow{1}{*}{$a_2=0.4$}   \\
                                                        &  \multirow{1}{*}{$a_3=0.1$}   \\
\hline
\multirow{3}{*}{The targeted data rates for users } & \multirow{1}{*}{$R_{{1}}=0.6 $ BPCU}  \\
                                                & \multirow{1}{*}{$R_{{2}}=1.6$ BPCU}  \\
                                                & \multirow{1}{*}{$R_{{3}} = 2$ BPCU}  \\
\cline{1-2}
The distance from BS to IRS  &  $d_{sr}=0.5$  \\
\hline
\multirow{3}{*}{The distance from IRS to users}  & \multirow{1}{*}{$d_{r1}=0.5$ } \\
                                                     & \multirow{1}{*}{$d_{r2}=0.4$  }  \\
                                                     & \multirow{1}{*}{$d_{r3}=0.3$  }  \\
\cline{1-2}
\end{tabular}
\label{parameter}
\end{table}
\section{Numerical Results}\label{Numerical Results}
In this section, the numerical results are presented to confirm the rationality of the derived theoretical expressions for IRS-NOMA networks. We show the impact of the reflecting elements on the performance of the IRS-NOMA communication network.  Monte Carlo simulation parameters used are summarized in TABLE~\ref{parameter}, where BPCU denotes the short for bit per channel use. Assume that three users $M=3$ are taken into consideration and the distance from the BS to IRS, and then to the terminal users are normalized to unity \cite{Men2017,Yue8026173}. Different from the channel gains modeled using the 3GPP Urban Micro in \cite{Emil2019}, the variances of complex channel coefficients are set to be ${\Omega _{sr}} = d_{sr}^{ - \alpha }$, ${\Omega _{r1}} = d_{r1}^{ - \alpha }$, ${\Omega _{r2}} = d_{r2}^{ - \alpha }$ and ${\Omega _{r3}} = d_{r3}^{ - \alpha }$, respectively. The complexity-vs-accuracy tradeoff parameter is set to be $N=20$ and simulation results are denoted by $ \bullet $. Without loss of the generality, the IRS-OMA and conventional OMA (i.e.,  variable gain AF relaying, FD/HD DF relaying) are selected as the benchmarks for the purpose of comparison. Here AF relaying works in HD mode and is equipped with a single antenna. The FD DF relaying is equipped with a pair of transceiver antennas, while HD DF relaying has a single antenna. The target rate ${R_{oma}}$ of the orthogonal user is equal to $\mathop \sum \limits_{i = 1}^M {R _i} $.

\begin{figure}[t!]
    \begin{center}
        \includegraphics[width=3.48in, height=2.8in]{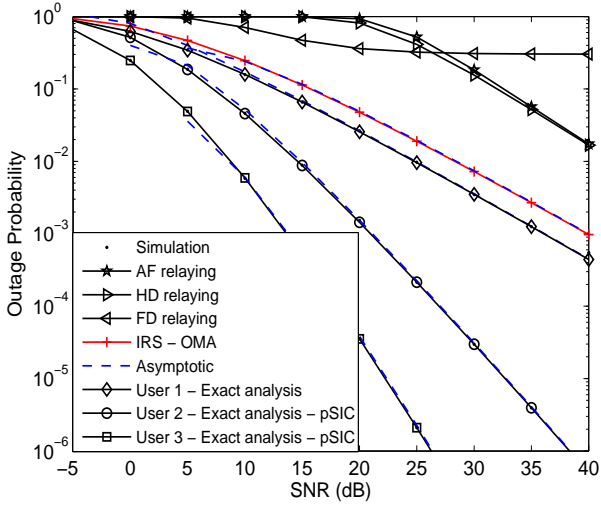}
        \caption{Outage probability versus the transmit SNR, with $K=1$, $Q=1$, $P=1$, $R_1=0.6$, $R_2=1.6$, $R_3=2$ and $R_{oma}=4.2$ BPCU. }
        \label{IRS_NOMA_OP_K1_VS_Relaying}
    \end{center}
\end{figure}

\subsection{Outage Probability}
Fig. \ref{IRS_NOMA_OP_K1_VS_Relaying} plots the outage probability of three users versus SNR for a simulation setting with $K=1$, $Q=1$, $P=1$,  $R_1=0.6$, $R_2=1.6$, $R_3=2$ and $R_{oma}=4.2$ BPCU. The theoretical analysis curves of outage probability for users with pSIC are plotted according to \eqref{the OP of the m-th user with pSIC under Rayleigh fading channel}. It is obvious that the Monte Carlo simulation outage probability curves excellently agree with analytical results across the entire average SNR range. The asymptotic outage probability converges to the analytical expressions given in \eqref{the diversity order of the m-th user with pSIC under Rayleigh fading channel}, which proves the effectiveness of our theoretical derivation. As can be seen from the figure that the outage performance of the nearest user ($M=3$) is higher than that of the distant users ($m=2$ and $m=1$). This is due to the fact that the nearby user  attains the higher diversity order, which verifies the insights in \textbf{Remark \ref{Remark2:the diversity order of the m-th user with pSIC under Rayleigh fading channel Q1}}. The exact and asymptotic outage probability cures of IRS-OMA are plotted according to the analytical results in \eqref{outage of OMA} and \eqref{the diversity order of OMA for Q1}, respectively.
One can observe that the outage behaviors of IRS-NOMA with pSIC are superior than that of IRS-OMA \eqref{outage of OMA}, variable gain AF relaying \cite{Laneman2004Cooperative}, FD relaying ~\cite[Eq. (7)]{Kwon2010Optimal} with loop self-interference i.e., $\mathbb{E}\{|h_{LI}|^2\}= -10$ dB and HD relaying  ~\cite[Eq. (8)]{Kwon2010Optimal}. The reasons are that: 1) IRS-NOMA can realize much better user fairness than IRS-OMA for multiple users; 2) FD DF relay suffers from loop interference due to signal leakage and needs the advanced loop interference cancellation technologies, which will lead to the higher cost; and 3) IRS-NOMA operates in FD mode provides the more spectrum efficient than HD DF relaying.

\begin{figure}[t!]
    \begin{center}
        \includegraphics[width=3.48in, height=2.8in]{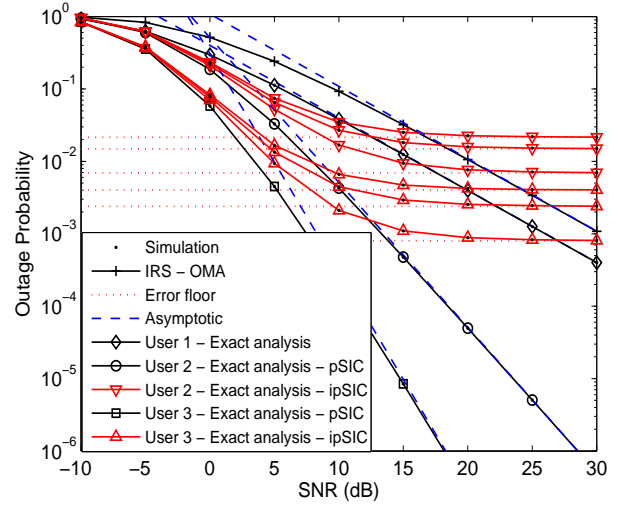}
        \caption{Outage probability versus the transmit SNR with different residual interference, $K = 2$, $Q=2$, $P=1$,  $R_1=0.6$, $R_2=1.6$, $R_3=2$ and $R_{oma}=4.2$ BPCU.}
        \label{IRS_NOMA_OP_K2_P1_Q2_diff_LI}
    \end{center}
\end{figure}

Fig. \ref{IRS_NOMA_OP_K2_P1_Q2_diff_LI} plots the outage probability of three users versus SNR for a simulation setting with $K = 2$, $Q=2$, $P=1$, $\varpi=1$, $R_1=0.6$, $R_2=1.6$, $R_3=2$ and $R_{oma}=4.2$ BPCU. The exact and approximate analyses curves of outage probability for users with ipSIC are plotted by \eqref{the OP of the m-th user with ipSIC under Rayleigh fading channel} and \eqref{the diversity order of the m-th user with ipSIC under Rayleigh fading channel}, respectively. The exact and asymptotic outage probability curves of IRS-OMA are plotted by  \eqref{outage of OMA} and \eqref{the diversity order of OMA for Q2}, respectively. The simulation results matches closely with the theoretical analysis. The important observation is that the outage probability of distant users with ipSIC converges to an error floor in the high SNR regime and thus obtain a zero diversity order. The reason is that there is the residual interference from ipSIC for IRS-NOMA. This phenomenon is also confirmed by the conclusions in \textbf{Remark \ref{Remark1:the diversity order of the m-th user with ipSIC under Rayleigh fading channel}}. Additionally, it is worth noting that the farthest user ($m=1$) does not carry out the SIC operation, since it has the worst channel conditions. Compared to the benchmark of IRS-OMA, we observe that IRS-NOMA with ipSIC is also capable of achieving the lower outage behaviors. Certainly, with the value of residual interference increasing, the achieved outage probability of IRS-NOMA converges to the worst error floors. As a result, it is important to consider the influence of ipSIC on the network performance for IRS-NOMA in the practical scenario.

\begin{figure}[t!]
    \begin{center}
        \includegraphics[width=3.48in, height=2.8in]{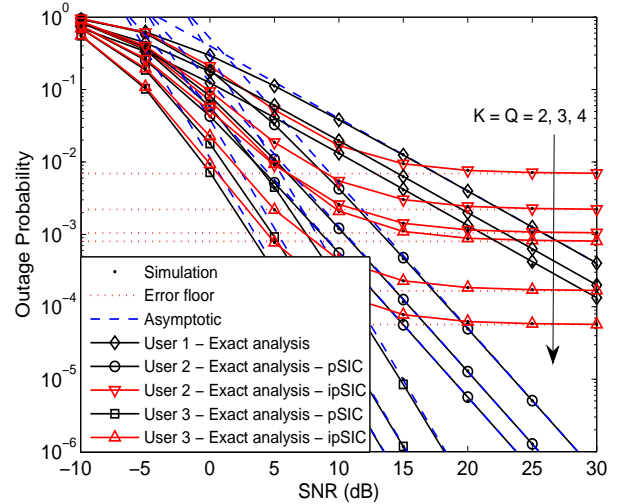}
        \caption{Outage probability versus the transmit SNR, with $P = 1$, $\varpi=1$, $R_1=0.6$, $R_2=1.6$, $R_3=2$ BPCU and $\mathbb{E}\{|h_{I}|^2\}=-10$ dB.}
        \label{IRS_NOMA_OP_fix_P1_diff_K_Q432}
    \end{center}
\end{figure}
\begin{figure}[t!]
    \begin{center}
        \includegraphics[width=3.48in, height=2.8in]{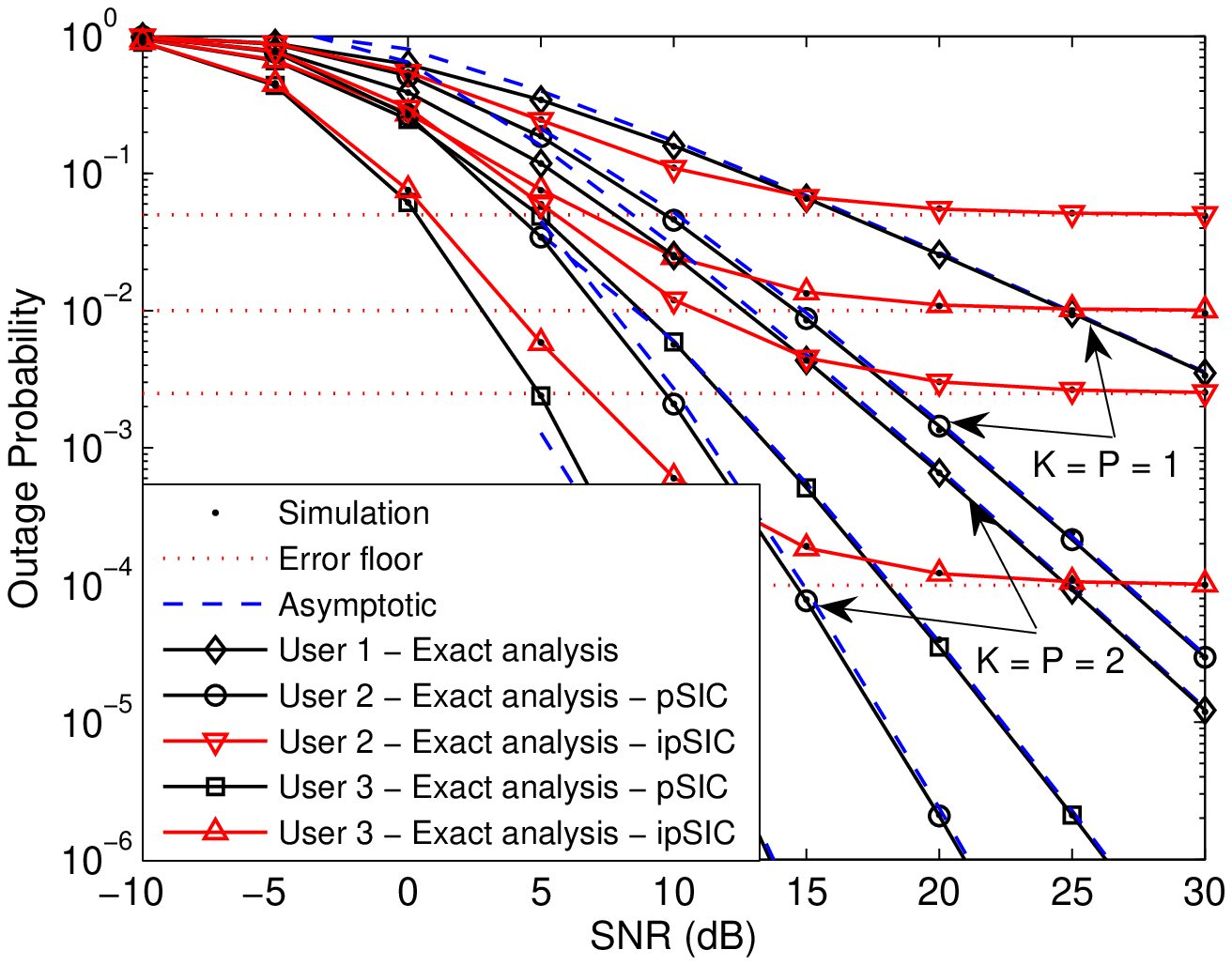}
        \caption{Outage probability versus the transmit SNR, with $\varpi=1$, $Q=1$, $R_1=0.6$, $R_2=1.6$, $R_3=2$ BPCU and $\mathbb{E}\{|h_{I}|^2\}=-10$ dB.}
        \label{IRS_NOMA_OP_fix_Q1_diff_K_P21}
    \end{center}
\end{figure}
\begin{figure}[t!]
    \begin{center}
        \includegraphics[width=3.48in, height=2.8in]{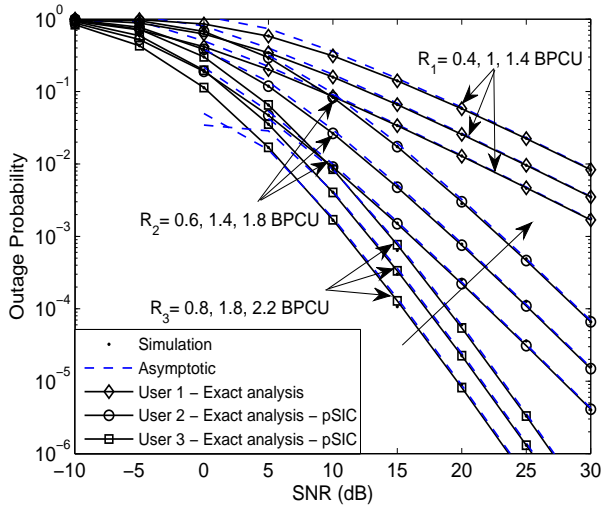}
        \caption{Outage probability versus the transmit SNR, with the different target rate for $K=2$, $P=2$ and $Q=1$.}
        \label{IRS_NOMA_OP_Q1_K2_P2_diff_rate}
    \end{center}
\end{figure}

Fig. \ref{IRS_NOMA_OP_fix_P1_diff_K_Q432} plots the outage probability versus SNR for a simulation system with different reflecting elements of IRS and $\mathbb{E}\{|h_{I}|^2\}=-10$ dB. One can observe that the setting of the reflecting elements for IRS-NOMA is significant to provide the network performance. With increasing the number of reflecting elements $K$, the lower outage probabilities are attained for multiple users.  These behaviors are caused by the fact that the application of IRS to NOMA networks provides a new degree of freedom to enhance the wireless link performance. This phenomenon is also certificate the completion of \textbf{Remark \ref{Remark2:the diversity order of the m-th user with pSIC under Rayleigh fading channel Q1}}, where both the number of reflecting elements and channel ordering determine the slope of outage probability for IRS-NOMA. Another observation is that all outage probability curves of each user have the same slopes, which manifests that the diversity orders of users are the same. This appearance demonstrates the insight we derived from the analytical results given by \eqref{the diversity order of the m-th user with pSIC under Rayleigh fading channel}.
\begin{figure}[t!]
    \begin{center}
        \includegraphics[width=3.48in, height=2.8in]{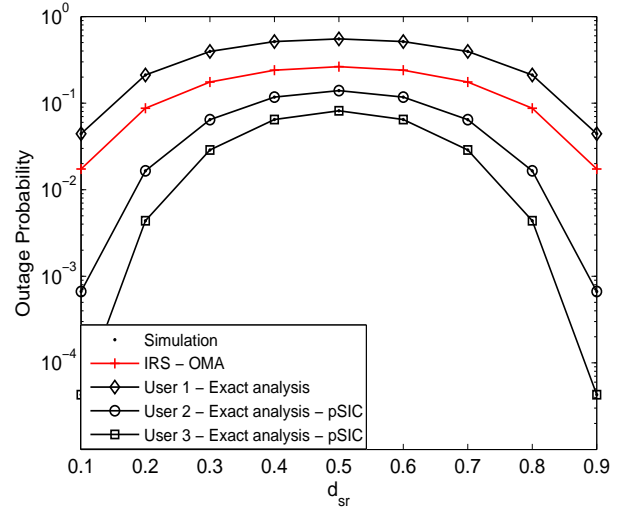}
        \caption{Outage probability versus distance $d_{sr}$ between the BS and IRS, with $Q=2$, $K=2$, $P=1$,  $R_1=R_2=R_3=0.6$ BPCU, $R_{oma}=1.8$ BPCU.}
        \label{IRS_NOMA_diff_distance}
    \end{center}
\end{figure}
\begin{figure}[t!]
    \begin{center}
        \includegraphics[width=3.48in, height=2.8in]{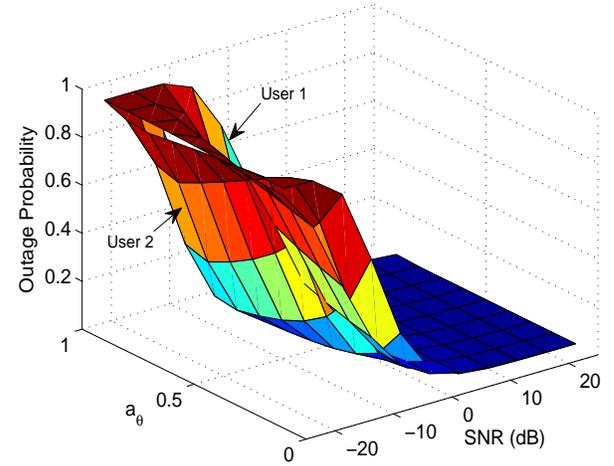}
        \caption{Outage probability versus the transmit SNR and $a_{\theta}$, with $K=1$, $P=1$, $Q=1$,  $R_1=0.1$ and $R_2=0.4$ BPCU.}
        \label{OP_3D}
    \end{center}
\end{figure}

Fig. \ref{IRS_NOMA_OP_fix_Q1_diff_K_P21} plots the outage probability versus SNR for a simulation setting with $Q=1$, $\varpi=1$, $R_1=0.6$, $R_2=1.6$, $R_3=2$ BPCU and $\mathbb{E}\{|h_{I}|^2\}=-10$ dB. The approximated outage probability curves of users are plotted corresponding to \eqref{the diversity order of the m-th user with pSIC under Rayleigh fading channel for Q1}, which match precisely with the simulation results. As can be observed from the figure that as the number of reflecting elements increases, the outage probability of users is becoming much smaller. The main reason behind this is that IRS-NOMA with 1-bit coding provides much more diversity orders given by \textbf{Remark \ref{Remark2:the diversity order of the m-th user with pSIC under Rayleigh fading channel Q1}}. It is worth mentioning that the outage probability curves of each user has a different diversity order, which confirms the analytical result derived in \eqref{the diversity order of the m-th user with pSIC under Rayleigh fading channel for Q1}. Fig. \ref{IRS_NOMA_OP_Q1_K2_P2_diff_rate} plots the outage probability versus SNR with the different target rate for $K=2$, $P=2$ and $Q=1$. One can observe that adjusting the target rate of users largely affect the outage performance. With the values of target rate increasing, the outage behaviors of users for IRS-NOMA networks are becoming much worse, which is in line with the conventional NOMA networks \cite{Yue8370069Unified}.

Fig. \ref{IRS_NOMA_diff_distance} plots the outage probability as a function of the normalized distance between the BS and users, with $Q=2$, $K=2$, $P=1$,  $R_1=R_2=R_3=0.6$ BPCU, $R_{oma}=1.8$ BPCU. We can observe that when the IRS is deployed closely to BS, the outage performance of non-orthogonal users is becoming much better. This phenomenon can be explained that the IRS can receive the clear LoS signals from the BS for the purpose of maximizing its received signal power. As the IRS departs from BS, the LoS deteriorates and outage probability of users increases seriously. When the IRS is in the middle of the BS and users, the worst outage behaviors of users are attained in IRS-NOMA networks. This is due to the fact that the IRS is neither closed to the BS nor to users. After this point, the performance begins to improve again. This is because that the IRS is close to NOMA users and enhance the reflecting signals received by users. Such an outage behavior can be useful to establish an optimal deployment of IRS in NOMA networks. As can be seen that the deployment scenarios of IRS should take into account some practical constraints.

To illustrate the impact of power allocation coefficients on system performance, Fig. \ref{OP_3D} plots the outage probability versus SNR and $a_{\theta}$, with with $K=1$, $P=1$, $Q=1$,  $R_1=0.1$ and $R_2=0.4$ BPCU, where ${a_\theta }$ is dynamic power allocation coefficients and the value range is zero to one i.e., ${a_\theta } \in \left[ {0,1} \right]$. We assume that there are a pair of users ($M=2$) in IRS-NOMA networks and the power coefficients of user 1 and user 2 have the relationships of $a_1=1-{a_\theta } $ and $a_2={a_\theta }$. The analytical curves of outage probability are plotted according to \eqref{the OP of the m-th user with pSIC under Rayleigh fading channel}. It is observed from the figure that with the value of  ${a_\theta }$ increasing, the outage behavior of user 1 deteriorates gradually, while the performance of user 2 first becomes better and then tends to worse. The reason for this phenomenon is that user 1 suffers more interference from user 2 when it detects its own information. At this time, user 2 needs to detect the signal of user 1 before detecting its own signal. Hence it is critical to design the power allocation coefficients for balancing the performance of two users.

\begin{figure}[t!]
    \begin{center}
        \includegraphics[width=3.48in, height=2.8in]{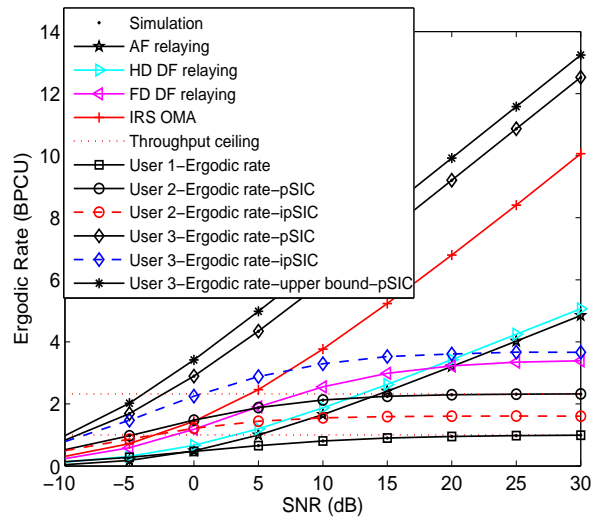}
        \caption{Rates versus the transmit SNR, with $K=1$, $Q=1$ and $P=1$.}
        \label{IRS_NOMA_Ergodic_Rate}
    \end{center}
\end{figure}
\begin{figure}[t!]
    \begin{center}
        \includegraphics[width=3.48in, height=2.8in]{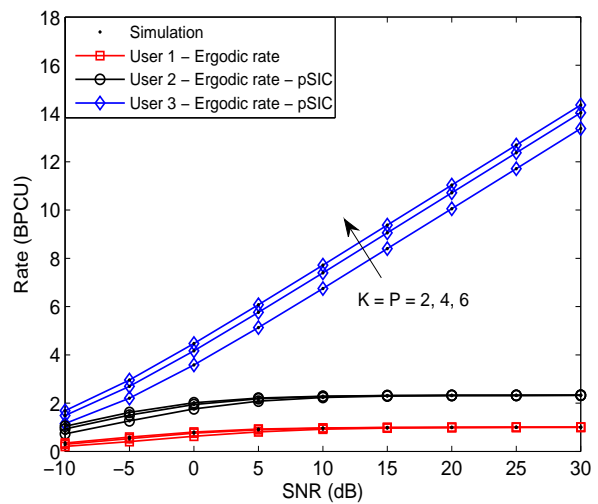}
        \caption{Rates versus the transmit SNR, with $Q=1$. }
        \label{IRS_NOMA_Ergodic_Rate_fix_Q1_diff_K_P}
    \end{center}
\end{figure}
\subsection{Ergodic Rate}
Fig. \ref{IRS_NOMA_Ergodic_Rate} plots the ergodic rates versus SNR, with $K=1$, $Q=1$ and $P=1$. The red and blue dotted curves denote the ergodic rates of the $m$-th user $(m=2)$ and $M$-th user $(M=3)$ with ipSIC for IRS-NOMA networks, which are plotted according to \eqref{the ergodic rate of the m-th user ipSIC} and \eqref{the ergodic rate of the M-th user ipSIC},  respectively. The exact curves of ergodic rate for the $m$-th and $M$-th user with pSIC are plotted based on \eqref{Ergodic Rate of the m-th user} and \eqref{Ergodic Rate of the M-th user}, respectively. One can observe that the ergdic rates of the $m$-th and $M$-th user with ipSIC are inferior to that of the $m$-th and $M$-th user with pSIC. The main reason is that the ergdic rates suffer from the residual interference of ipSIC and tends to constant values at high SNRs. Another observation is that the ergodic rates of distant users for IRS-NOMA outperform that of AF relaying and FD/HD relaying in the low SNR regime, which are consistent to FD/HD NOMA systems\cite{Zhong7572025,Yue8026173}.
As the SNR value increases, the ergodic rate of distant users converges to a throughput ceiling, which is also confirmed in \textbf{Remark \ref{Remark3}}. This is due to the fact that the distant user will suffer from the interference from the nearby users' signals when it decodes their own signals. Another observation is that the ergodic rate of nearest user is much greater than that of non-orthogonal users, IRS-OMA \eqref{the ergodic rate for OMA}, AF relaying and FD/HD relaying. The origin for this behavior is that it is closest to the IRS and has the best channel conditions. In addition, Fig. \ref{IRS_NOMA_Ergodic_Rate_fix_Q1_diff_K_P} plots the ergodic rates versus SNR with different reflecting elements. As can be observed from this figure that with the increasing reflecting elements of IRS, the ergodic performance of the nearest user with pSIC is enhanced and has the same slopes, which confirms the insights in \textbf{Remark \ref{Remark4}}. The distant users' performance has no obvious variety due to the effects of interference signals. We conclude that IRS-NOMA cannot circumvent the problem of $zero$ slope for the distant users.

\subsection{Energy Efficiency}
\begin{figure}[t!]
    \begin{center}
        \includegraphics[width=3.48in, height=2.8in]{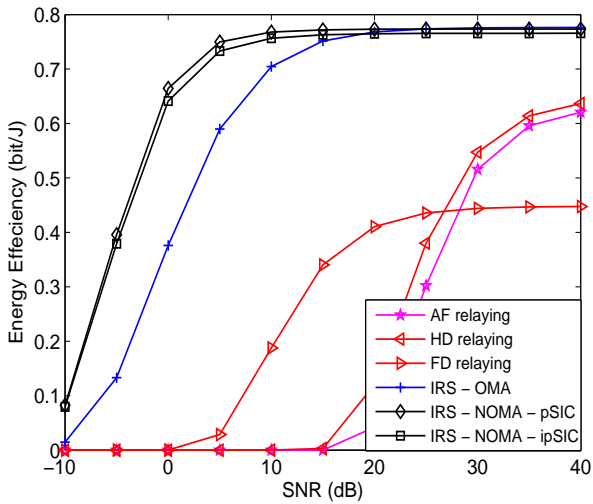}
        \caption{Energy efficiency in delay-limited transmission mode, where $K=2$, $Q=2$, $P=1$, $\kappa = 1.2$, $P_S $ = 5 dBW, $P_{BS} $ = 2 dBW, ${P_{m,UE}}$ = 10 dBm and ${P_k}\left( b \right)$ = 10 dBm.}
        \label{IRS_NOMA_EE_limited}
    \end{center}
\end{figure}
\begin{figure}[t!]
   \begin{center}
       \includegraphics[width=3.48in, height=2.8in]{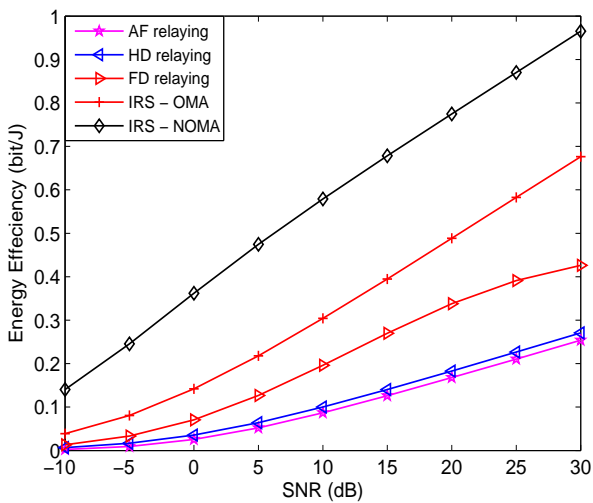}
       \caption{Energy efficiency in delay-tolerant transmission mode, where $K=2$, $Q=2$, $P=1$, $\kappa = 1.5$,  $P_S $ = 10 dBW, $P_{BS} $ = 4 dBW, ${P_{m,UE}}$ = 10 dBm and ${P_k}\left( b \right)$ = 10 dBm.}
       \label{IRS_NOMA_EE_tolerant}
   \end{center}
\end{figure}

Fig. \ref{IRS_NOMA_EE_limited} plots the energy efficiency for IRS-NOMA networks in the delay-limited transmission mode, with $K=2$, $Q=2$, $P=1$, $\kappa = 1.2$, $P_S $ = 5 dBW, $P_{BS} $ = 2 dBW, ${P_{m,UE}}$ = 10 dBm and ${P_k}\left( b \right)$ = 10 dBm. The energy efficiency curve of IRS-NOMA networks with ipSIC/pSIC is plotted according to \eqref{The Energy Efficiency of IRS-NOMA}. It can be observed that the energy efficiency of IRS-NOMA outperforms that of IRS-OMA and converge to the same value in the high SNR regime.  This is due to the fact that the setting of  sum target rate for non-orthogonal users is equal to the orthogonal user in delay-limited transmission mode. Another observation is that the energy efficiencies of IRS-NOMA and IRS-OMA are superior to that of AF relaying and FD/HD DF relaying. The main reason is that the IRS-assisted wireless communications are capable of improving the energy efficiency compared to these conventional cooperative communications. As a further advance, Fig. \ref{IRS_NOMA_EE_tolerant} plots the energy efficiency for IRS-NOMA in delay-tolerant transmission mode, with $K=2$, $Q=2$, $P=1$, $\kappa = 1.5$,  $P_S $ = 10 dBW, $P_{BS} $ = 4 dBW, ${P_{m,UE}}$ = 10 dBm and ${P_k}\left( b \right)$ = 10 dBm. The energy efficiency curve of IRS-NOMA networks with pSIC is plotted according to \eqref{The Energy Efficiency of IRS-NOMA}. We can observe that the energy efficiency of IRS-NOMA is much larger than that of IRS-OMA, AF relaying, FD/HD DF relaying. This is due to that IRS-NOMA with pSIC is able to achieve the larger system throughput relative to these benchmarks.
\section{Conclusion}\label{Conclusion}
In this paper, an IRS has been invoked in downlink NOMA networks for enhancing the performance of multiple users, where 1-bit coding scheme is taken into account. More specifically, we have derived the exact expressions of outage probability and ergodic rate for users with ipSIC/pSIC in IRS-assisted NOMA networks. Based on the approximated analyses, the diversity order of the $m$-th user is related to the number of reflecting elements and channel ordering. With increasing the number of reflecting elements, the outage probability of users with ipSIC/pSIC for IRS-NOMA networks decreases. Due to the interference from the nearby users' signals, a $zero$ of high SNR slope for ergodic rate is obtained by the distant users. Simulation results have shown that the outage behaviors of IRS-NOMA are superior to that of IRS-OMA, AF relaying and FD/HD DF relaying. The nearest user has a larger ergodic rate than orthogonal user and distant users. Finally, the throughput and energy efficiency of IRS-NOMA were discussed both in delay-limited and delay-tolerant transmission modes. The application of IRS to NOMA provided a new degree of freedom to enhance the wireless link performance.

\appendices
\section*{Appendix~A: Proof of Theorem \ref{Theorem1:the OP of the m-th user with ipSIC under Rayleigh fading channel}} \label{Appendix:A}
\renewcommand{\theequation}{A.\arabic{equation}}
\setcounter{equation}{0}
The proof starts by assuming ${W_p} = {\rm{min}}\left( {{{\rm{E}}_{m,1}},{{\rm{E}}_{m,2}}, \cdots ,{{\rm{E}}_{m,m}}} \right)$, $W = \max \left( {{W_1},{W_2}...,{W_P}} \right)$ and then
\begin{align}\label{AppendixA: P_m}
{P_m} = \Pr \left( W \right) &= \Pr \left[ {\max \left( {{W_1},{W_2}...,{W_P}} \right)} \right] \nonumber \\
  &= \prod\limits_{p = 1}^P {\Pr \left( {{W_p}} \right)}  = {\left[ {\Pr \left( {{W_p}} \right)} \right]^P} .
\end{align}
Hence the outage probability of the $m$-th user with ipSIC need to further calculate $\Pr \left( {{W_p}} \right)$. Applying the complementary set and some algebraic manipulations, it can be calculated as follows:
\begin{align}\label{AppendixA: OP expression of the m-th user with Rice fading}
&\Pr \left( {{W_p}} \right) = {\rm{Pr}}\left[ {{{\left| {{\bf{v}}_p^H{{\bf{D}}_m}{{\bf{h}}_{sr}}} \right|}^2} < \psi _m^ * \left( {\varpi \rho {{\left| {{h_I}} \right|}^2} + 1} \right)} \right] \nonumber \\
& = \int_0^\infty  {{F_{{{\left| {{\bf{v}}_p^H{{\bf{D}}_m}{ {\bf{ h_{sr}}}}} \right|}^2}}}\left( {\psi _m^ * \left( {\varpi \rho y + 1} \right)} \right){f_{{{\left| {{h_I}} \right|}^2}}}\left( y \right)dy},
\end{align}
where $\varpi=1$, ${f_{{{\left| {{h_I}} \right|}^2}}}\left( y \right) = \frac{1}{{{\Omega _I}}}{e^{ - \frac{y}{{{\Omega _I}}}}}$, $\psi _m^ *  = \max \left\{ {{\psi _1}, \ldots ,{\psi _m}} \right\}$, ${\psi _m} = \frac{{{\gamma _{t{h_m}}}}}{{\rho \left( {{a_m} - {\gamma _{t{h_m}}}\sum\nolimits_{i = m + 1}^M {{a_i}} } \right)}}$ with ${a_m} > {\gamma _{t{h_m}}}\sum\nolimits_{i = m + 1}^M {{a_i}} $ and ${\psi _M} = \frac{{{\gamma _{t{h_M}}}}}{{\rho {a_M}}}$. Based on the previous assumption, the cascade channel gains from the BS to IRS and then to users with 1-bit coding scheme are also sorted as ${\left| {{\bf{v}}_p^H{{\bf{D}}_1}{{\bf{h}}_{sr}}} \right|^2} \le \cdots  \le {\left| {{\bf{v}}_p^H{{\bf{D}}_m}{{\bf{h}}_{sr}}} \right|^2} \le  \cdots  \le {\left| {{\bf{v}}_p^H{{\bf{D}}_M}{{\bf{h}}_{sr}}} \right|^2}$. As a further advance, we focus our attention the CDF ${F_{{{\left| {{\bf{v}}_p^H{{\bf{D}}_m}{{\bf{h}}_{sr}}} \right|}^2}}}\left( x \right)$ in the following part.

Denoting $X = {\left| {{\bf{v}}_p^H{{\bf{D}}_m}{{\bf{h}}_{sr}}} \right|^2}$, the CDF ${F_{X}}\left( x \right)$ of the cascade channel gain has a fixed relationship with the unsorted channel gain \cite{David2003Order,Yue8370069Unified}, which can be expressed as
\begin{align}\label{AppendixA:the relationship with the unsorted channel gain}
{F_X}\left( x \right) =& \frac{{M!}}{{\left( {M - m} \right)!\left( {m - 1} \right)!}}\sum\limits_{l = 0}^{M - m} {{
   {M - m}  \choose
   l  }}  \nonumber \\
  &\times \frac{{{{\left( { - 1} \right)}^l}}}{{m + l}} {\left[ {{\bar F_{ X}}\left( x \right)} \right]^{m + l}} ,
\end{align}
where ${{{\bar F}_X}\left( x \right)}$ denotes the CDF of unsorted cascade channel gain. After some arithmetical manipulations, we can further express $X = {\left| {\sum\nolimits_{k = 1}^Q {h_{sr}^kh_{rm}^k} } \right|^2}$, where $h_{sr}^k \sim {\cal C}{\cal N}\left( {0,{\Omega _{sr}}} \right)$ and $h_{rm}^k \sim {\cal C}{\cal N}\left( {0,{\Omega _{rm}}} \right)$. Based on \cite[Eq. (7)]{Liu2014cascade}, the PDF ${{\bar f}_X}\left( x \right)$ of the cascade channel gain $X={\left| {{\bf{v}}_p^H{{\bf{D}}_m}{{\bf{h}}_{sr}}} \right|^2}$ from the BS to IRS and then to users is given by
\begin{align}\label{AppendixB: the PDF of cascade channels for unsorted channel}
{{\bar f}_X}\left( x \right) = \frac{{2{x^{\frac{{Q - 1}}{2}}}}}{{\Gamma \left( Q \right){{\left( {\sqrt {{\Omega _{sr}}{\Omega _{rm}}} } \right)}^{Q + 1}}}}{K_{Q - 1}}\left( {2\sqrt {\frac{x}{{{\Omega _{sr}}{\Omega _{rm}}}}} } \right),
\end{align}
where ${{{\bar f}_X}\left( x \right)}$ denotes the PDF of unsorted cascade channel gain. Furthermore, we perform integral processing on the above formula and express the corresponding CDF ${{\bar F}_X}\left( x \right)$ as
\begin{align}\label{AppendixB: the CDF of cascade channels for unsorted channel}
{{\bar F}_X}\left( x \right) = \frac{2}{\Upsilon }\int_0^x {{y^{\frac{{Q - 1}}{2}}}} {K_{Q - 1}}\left( {2\sqrt {\frac{y}{{{\Omega _{sr}}{\Omega _{rm}}}}} } \right)dy,
\end{align}
where $\Upsilon  = \Gamma \left( Q \right){\left( {\sqrt {{\Omega _{sr}}{\Omega _{rm}}} } \right)^{Q + 1}}$.

Based on \eqref{AppendixB: the CDF of cascade channels for unsorted channel}, using $y = xt$ and \cite[Eq. (6.561.8)]{2000gradshteyn}, $ {{\bar F}_X}\left( x \right) $ can be calculated as follows:
\begin{align}\label{AppendixB: the CDF details of cascade channels for unsorted channel}
{{\bar F}_X}\left( x \right) &= \frac{2}{\Upsilon }\int_0^1 {{{\left( {\sqrt {xt} } \right)}^{Q - 1}}} {K_{Q - 1}}\left( {2\sqrt {\frac{{xt}}{{{\Omega _{sr}}{\Omega _{rm}}}}} } \right)xdt \nonumber \\
& \mathop  = \limits^{\sqrt t  \buildrel \Delta \over = y} \frac{{4{x^{\frac{{Q + 1}}{2}}}}}{\Upsilon }\int_0^1 {{y^Q}{K_{Q - 1}}\left( {2\sqrt {\frac{x}{{{\Omega _{sr}}{\Omega _{rm}}}}} y} \right)dy} \nonumber \\
 &= 1 - \frac{2}{{\Gamma \left( Q \right)}}{\left( {\frac{x}{{{\Omega _{sr}}{\Omega _{rm}}}}} \right)^{\frac{Q}{2}}}{K_Q}\left( {2\sqrt {\frac{x}{{{\Omega _{sr}}{\Omega _{rm}}}}} } \right).
\end{align}
Upon substituting \eqref{AppendixB: the CDF details of cascade channels for unsorted channel} into \eqref{AppendixA:the relationship with the unsorted channel gain}, the CDF ${F_{{{X }}}}\left( x \right)$ of the cascade channel gain can be given by
\begin{align}\label{AppendixB: the CDF of cascade channels for sorted channel}
{F_X}\left( x \right) =& {\phi _m}\sum\limits_{l = 0}^{M - m} {{
   {M - m}  \choose
   l }} \frac{{{{\left( { - 1} \right)}^l}}}{{m + l}}\left[ {1 - \frac{2}{{\Gamma \left( Q \right)}}} \right.\nonumber \\
  & \times {\left. {{{\left( {\frac{x}{{{\Omega _{sr}}{\Omega _{rm}}}}} \right)}^{\frac{Q}{2}}}{K_Q}\left( {2\sqrt {\frac{x}{{{\Omega _{sr}}{\Omega _{rm}}}}} } \right)} \right]^{m + l}} ,
\end{align}
where ${\phi _m} = \frac{{M!}}{{\left( {M - m} \right)!\left( {m - 1} \right)!}}$.

Combining \eqref{AppendixB: the CDF of cascade channels for sorted channel} and \eqref{AppendixA: OP expression of the m-th user with Rice fading}, the probability  $\Pr \left( {{W_p}} \right)$ can be further expressed as follows:

\begin{align}\label{AppendixA: PrW_p}
 &\Pr \left( {{W_p}} \right) = \frac{{{\phi _m}}}{{{\Omega _I}}}\sum\limits_{l = 0}^{M - m} {{
   {M - m}  \choose
   l  }} \frac{{{{\left( { - 1} \right)}^l}}}{{m + l}}\int_0^\infty  {\left[ {1 - \frac{2}{{\Gamma \left( Q \right)}}} \right.}  \nonumber \\
 &{\left. { \times {{\left( {\frac{{\psi _m^ * \varphi }}{{{\Omega _{sr}}{\Omega _{rm}}}}} \right)}^{\frac{Q}{2}}}{K_Q}\left( {2\sqrt {\frac{{\psi _m^ * \varphi }}{{{\Omega _{sr}}{\Omega _{rm}}}}} } \right)} \right]^{m + l}}{e^{ - \frac{y}{{{\Omega _I}}}}}dy ,
\end{align}
where $\varphi  = \left( {\varpi \rho y + 1} \right)$. Assuming $x = \frac{y}{{{\Omega _I}}}$, and using Gauss-Laguerre integration \cite[Eq. (25.4.45)]{Abramowitz1972Handbook}, the probability $\Pr \left( {{W_p}} \right)$ can be given by
\begin{align}\label{AppendixB: PrW_p}
\Pr \left( {{W_p}} \right) \approx &  {\phi _m}\sum\limits_{l = 0}^{M - m} {\sum\limits_{u = 1}^U {{
   {M - m}  \choose
   l  }} } \frac{{{{\left( { - 1} \right)}^l}{H_u}}}{{m + l}}\left[ {1 - \frac{2}{{\Gamma \left( Q \right)}}} \right. \nonumber \\
& {\left. { \times {{\left( {\frac{{\psi _m^ * \Lambda}}{{{\Omega _{sr}}{\Omega _{rm}}}}} \right)}^{\frac{Q}{2}}}{K_Q}\left( {2\sqrt {\frac{{\psi _m^ * \Lambda}}{{{\Omega _{sr}}{\Omega _{rm}}}}} } \right)} \right]^{m + l}} ,
\end{align}
where $\Lambda  = \left( {\varpi \rho {\Omega _I}{x_u} + 1} \right)$, ${H_u}$ is the weight of the Gauss-Laguerre integration. ${{{ r}_u}}$ is the $u$-th zero of Laguerre polynomial ${{ L}_U}\left( {{{ r}_u}} \right)$ and the corresponding the $u$-th weight is given by ${H_u} = \frac{{{{\left( {U!} \right)}^2}{r_u}}}{{{{\left[ {{L_{U + 1}}\left( {{r_u}} \right)} \right]}^2}}}$.

Upon substituting \eqref{AppendixB: PrW_p} into \eqref{AppendixA: P_m}, we can obtain \eqref{the OP of the m-th user with ipSIC under Rayleigh fading channel}. The proof is completed.

\appendices
\section*{Appendix~B: Proof of Theorem \ref{Theorem:Ergodic Rate of the m-th user}} \label{Appendix:B}
\renewcommand{\theequation}{B.\arabic{equation}}
\setcounter{equation}{0}

Upon substituting $\varpi  = 0$ into \eqref{the ergodic rate of the m-th user ipSIC}, the ergodic rate of the $m$-th user with pSIC for IRS-NOMA networks can be written as
\begin{align}\label{AppendixC:The ergodic rate expression with pSIC}
R_{m,erg}^{pSIC}  &= {\mathbb{E}}\left\{ {\log \left( {1 + \underbrace {\mathop {\max }\limits_{{{\bf{v}}_p}} \frac{{\rho {{\left| {{\bf{v}}_p^H{{\bf{D}}_m}{{\bf{h}}_{sr}}} \right|}^2}{a_m}}}{{\rho {{\left| {{\bf{v}}_p^H{{\bf{D}}_m}{{\bf{h}}_{sr}}} \right|}^2}{{\bar a}_m} + 1}}}_Y} \right)} \right\}  \nonumber \\
&  = \frac{1}{{\ln 2}}\int_0^\infty  {\frac{{1 - {F_Y}\left( y \right)}}{{1 + y}}dy}.
\end{align}

Let $\bar Y = \frac{{\rho {{\left| {{\bf{v}}_p^H{{\bf{D}}_m}{{\bf{h}}_{sr}}} \right|}^2}{a_m}}}{{\rho {{\left| {{\bf{v}}_p^H{{\bf{D}}_m}{{\bf{h}}_{sr}}} \right|}^2}{{\bar a}_m} + 1}}$ and the CDF ${F_{\bar Y}}\left( y \right)$ can be expressed as
\begin{align}\label{AppendixC:The ergodic rate expression with pSIC detail}
{F_{\bar Y}}\left( y \right) = \Pr \left[ {{{\left| {{\bf{v}}_p^H{{\bf{D}}_m}{{\bf{h}}_{sr}}} \right|}^2} < \frac{y}{{\rho \left( {{a_m} - y{{\bar a}_m}} \right)}}} \right],
\end{align}
where ${a_m} > y{{\bar a}_m}$. Upon substituting \eqref{AppendixB: the CDF of cascade channels for sorted channel} into \eqref{AppendixC:The ergodic rate expression with pSIC detail} and using Binomial theorem, the CDF ${F_{\bar Y}}\left( y \right)$ is given by
\begin{align}\label{AppendixC:The ergodic rate expression with pSIC detail order}
& {F_{\bar Y}}\left( y \right) = {\phi _m}\sum\limits_{l = 0}^{M - m} {\sum\limits_{r = 0}^{m + l} {{
   {M - m}  \choose
   l  }
   } } {
   {m + l}  \choose
   r  }\frac{{{{\left( { - 1} \right)}^{l + r}}}}{{m + l}} \nonumber \\
 & \times{\left( {\frac{2}{{\Gamma \left( Q \right)}}} \right)^r} {\left[ {{{\left( {\frac{{y{\varphi _m}}}{{\left( {{a_m} - y{{\bar a}_m}} \right)}}} \right)}^{\frac{Q}{2}}}{K_Q}\left( {2\sqrt {\frac{{y{\varphi _m}}}{{\left( {{a_m} - y{{\bar a}_m}} \right)}}} } \right)} \right]^r},
\end{align}
where ${\varphi _m} = \frac{1}{{\rho {\Omega _{sr}}{\Omega _{rm}}}}$.

For IRS-NOMA with 1-bit coding scheme, ${{\bf{v}}_p^H{{\bf{D}}_m}{{\bf{h}}_{sr}}}$ and ${{\bf{v}}_l^H{{\bf{D}}_m}{{\bf{h}}_{sr}}}$ are independent and identically distribution for $p \ne l$. As a consequence, the CDF  ${F_Y}\left( y \right)$ can be given by
\begin{align}\label{AppendixC:The ergodic rate expression with pSIC detail order final}
{F_Y}\left( y \right) =& \left\{ {{\phi _m}\sum\limits_{l = 0}^{M - m} {\sum\limits_{r = 0}^{m + l} {{
   {M - m}  \choose
   l  }} } {
   {m + l}  \choose
   r  }} \right.  \nonumber \\
  &\times \frac{{{{\left( { - 1} \right)}^{l + r}}}}{{m + l}}\left[ {\frac{2}{{\Gamma \left( Q \right)}}{{\left( {\frac{{y{\varphi _m}}}{{\left( {{a_m} - y{{\bar a}_m}} \right)}}} \right)}^{\frac{Q}{2}}}} \right.\nonumber \\
 &{\left. { \times {{\left. {{K_Q}\left( {2\sqrt {\frac{{y{\varphi _m}}}{{\left( {{a_m} - y{{\bar a}_m}} \right)}}} } \right)} \right]}^r}} \right\}^P} .
\end{align}
Upon substituting \eqref{AppendixC:The ergodic rate expression with pSIC detail order final} into \eqref{AppendixC:The ergodic rate expression with pSIC}, the CDF  ${F_Y}\left( y \right)$ can be expressed as follows:
\begin{align}\label{AppendixC:The ergodic rate expression with pSIC detail order final with Gauss}
R_{m,erg}^{pSIC} =& \frac{1}{{\ln 2}}\int_0^{\frac{{{a_m}}}{{{{\bar a}_m}}}} {\frac{1}{{1 + y}}\left\langle {1 - \left[ {{\phi _m}\sum\limits_{l = 0}^{M - m} {\sum\limits_{r = 0}^{m + l} {{
   {M - m}  \choose
   l  }} } } \right.} \right.} \nonumber \\
   & \times {
   {m + l}  \choose
   r }\frac{{{{\left( { - 1} \right)}^{l + r}}}}{{m + l}}\left[ {\frac{2}{{\Gamma \left( Q \right)}}{{\left( {\frac{{y{\varphi _m} }}{{ \left( {{a_m} - y{{\bar a}_m}} \right)}}} \right)}^{\frac{Q}{2}}}} \right.\nonumber \\
   & \left. {{{\left. { \times {{\left. {{K_Q}\left( {2\sqrt {\frac{{y {\varphi _m}}}{{ \left( {{a_m} - y{{\bar a}_m}} \right)}}} } \right)} \right]}^r}} \right\}}^P}} \right\rangle dy.
\end{align}
By further applying the Gauss-Chebyshev quadrature \cite{Hildebrand1987introduction} on the above equation, we can obtain \eqref{Ergodic Rate of the m-th user}. The proof is completed.

\appendices
\section*{Appendix~C: Proof of Theorem \ref{Theorem:Ergodic Rate of the M-th user}} \label{Appendix:C}
\renewcommand{\theequation}{C.\arabic{equation}}
\setcounter{equation}{0}

Upon substituting $\varpi  = 0$ into \eqref{the ergodic rate of the M-th user ipSIC}, the ergodic rate of the $M$-th user with pSIC for IRS-NOMA networks can be written as
\begin{align}\label{AppendixC:The ergodic rate expression for M-th user with pSIC}
R_{M,erg}^{pSIC}& = {\mathbb{E}}\left[ {\log \left( {1 + \rho {a_M}\underbrace {\mathop {\max }\limits_{{{\bf{v}}_p}} \left( {{{\left| {{\bf{v}}_p^H{{\bf{D}}_M}{{\bf{h}}_{{{sr}}}}} \right|}^2}} \right)}_Z} \right)} \right] \nonumber \\
& = \frac{{\rho {a_M}}}{{\ln 2}}\int_0^\infty  {\frac{{1 - {F_Z}\left( z \right)}}{{1 + z\rho {a_M}}}dx} .
\end{align}

Denoting $\bar Z = {\left| {{\bf{v}}_p^H{{\bf{D}}_M}{{\bf{h}}_{{{sr}}}}} \right|^2}$, by the virtue of Order Statistic theory and \eqref{AppendixB: the CDF details of cascade channels for unsorted channel}, the CDF ${F_{\bar Z}}\left( z \right)$ of the $M$-th user is given by
\begin{align}\label{AppendixC:The ergodic rate expression for M-th user with pSIC}
{F_{\bar Z}}\left( z \right) = {\left[ {1 - \frac{2}{{\Gamma \left( Q \right)}}{{\left( {\frac{z}{{{\Omega _{sr}}{\Omega _{rM}}}}} \right)}^{\frac{Q}{2}}}{K_Q}\left( {2\sqrt {\frac{z}{{{\Omega _{sr}}{\Omega _{rM}}}}} } \right)} \right]^M}.
\end{align}
Similar to the procedure in \eqref{AppendixC:The ergodic rate expression with pSIC detail order final}, we randomly select a column ${{{\bf{v}}_p}}$ from $\bf{V} $ to maximize $\bar Z $. Hence the CDF ${F_Z}\left( z \right)$ can be given by
\begin{align}\label{AppendixC:The CDF for M-th user with pSIC}
{F_Z}\left( z \right) = &\left[ {1 - \frac{2}{{\Gamma \left( Q \right)}}} \right.  \nonumber \\
 &{\left. { \times {{\left( {\frac{z}{{{\Omega _{sr}}{\Omega _{rM}}}}} \right)}^{\frac{Q}{2}}}{K_Q}\left( {2\sqrt {\frac{z}{{{\Omega _{sr}}{\Omega _{rM}}}}} } \right)} \right]^{MP}}   .
\end{align}
Applying the Binomial theorem in \eqref{AppendixC:The CDF for M-th user with pSIC}, the CDF ${F_X}\left( x \right)$ can be further rewritten as follow:
\begin{align}\label{AppendixC:The CDF for M-th user with pSIC for Binomial}
 {F_Z}\left( z \right) =& \sum\limits_{r = 0}^{MP} {{
   {MP}  \choose
   r }} {\left( { - 1} \right)^{r + 1}}\left[ {\frac{2}{{\Gamma \left( Q \right)}}} \right. \nonumber \\
  &\times {\left. {{{\left( {\frac{z}{{{\Omega _{sr}}{\Omega _{rm}}}}} \right)}^{\frac{Q}{2}}}{K_Q}\left( {2\sqrt {\frac{z}{{{\Omega _{sr}}{\Omega _{rm}}}}} } \right)} \right]^r} .
\end{align}

By substituting \eqref{AppendixC:The CDF for M-th user with pSIC for Binomial} and after some manipulates, we can obtain \eqref{Ergodic Rate of the M-th user}. The proof is completed.

\bibliographystyle{IEEEtran}
\bibliography{mybib}

\end{document}